\documentclass[reprint,amsmath,amssymb,aps,pra,superscriptaddress,floatfix, nofootinbib]{revtex4-2}
\usepackage{hyperref}
\usepackage[utf8]{inputenc}
\usepackage[page]{appendix}
\usepackage[english]{babel}
\usepackage{float}
\usepackage{graphicx}
\usepackage[capitalise]{cleveref}
\usepackage[version=4]{mhchem}
\usepackage{tabularx}
\usepackage[dvipsnames]{xcolor}
\usepackage[per-mode=power, 
             exponent-product = \times, 
             inter-unit-product = \ensuremath{{\cdot}}, 
             tight-spacing = true,
             number-unit-product=~,
             parse-numbers=false,]{siunitx}
 \AtBeginDocument{\RenewCommandCopy\qty\SI}
\usepackage[
    protrusion=true,
    activate={true,nocompatibiality},
    final,
    tracking=true,
    kerning=true,
    spacing=true,
    factor=1000]{microtype}
\hypersetup{
 colorlinks=true, 
 linkcolor=blue, 
 citecolor=blue,        
 filecolor=blue,      
 urlcolor=blue}

\usepackage{caption}
\newcommand{\figsize}{0.98\textwidth}

\begin{document}
                                                         
\author{Gr\'egory Moille}
\email{gregory.moille@nist.gov}
\affiliation{Joint Quantum Institute, NIST/University of Maryland, College Park, USA}
\affiliation{Microsystems and Nanotechnology Division, National Institute of Standards and Technology, Gaithersburg, USA}
\author{Daron Westly}
\affiliation{Microsystems and Nanotechnology Division, National Institute of Standards and Technology, Gaithersburg, USA}
\author{Rahul Shrestha}
\affiliation{Joint Quantum Institute, NIST/University of Maryland, College Park, USA}
\author{Khoi Tuan Hoang}
\affiliation{Joint Quantum Institute, NIST/University of Maryland, College Park, USA}
\affiliation{Microsystems and Nanotechnology Division, National Institute of Standards and Technology, Gaithersburg, USA}
\author{Kartik Srinivasan}
\email{kartik.srinivasan@nist.gov}
\affiliation{Joint Quantum Institute, NIST/University of Maryland, College Park, USA}
\affiliation{Microsystems and Nanotechnology Division, National Institute of Standards and Technology, Gaithersburg, USA}
\date{\today}

\title{Broadband Visible Wavelength Microcomb Generation In Silicon Nitride Microrings Through Air-Clad Dispersion Engineering}
\begin{abstract}
The development of broadband microresonator frequency combs at visible wavelengths is pivotal for the advancement of compact and fieldable optical atomic clocks and spectroscopy systems. Yet, their realization necessitates resonators with anomalous dispersion, an arduous task due to the prevailing normal dispersion regime of materials within the visible spectrum. In this work, we evince that silicon nitride microring resonators with air cladding on top and sides -- a deviation from the frequently employed silica-embedded resonators -- allows for the direct generation of broadband microcombs in the visible range.  We experimentally demonstrate combs pumped at \qty{1060}{\nm} (\qty{283}{\THz}) that reach wavelengths as short as \qty{680}{\nm} (\qty{440}{\THz}), and combs pumped at \qty{780}{\nm} (\qty{384}{\THz}) that reach wavelengths as short as \qty{630}{\nm} (\qty{475}{\THz}). We further show through simulations that microcombs extending to wavelengths as low as \qty{461}{\nm} (\qty{650}{\THz}) should be accessible in this platform.    
\end{abstract}


\maketitle

Photonic integrated circuits, thanks to their scalability and manufacturability, have revolutionized the availability of critical optical functionalities in compact and deployable formats~\cite{ColeIEEECommun.Mag.2009,FahrenkopfIEEEJ.Sel.Top.QuantumElectron.2019,ThomsonJ.Opt.2016a}. Historically designed specifically for telecommunication wavelengths around \qty{1550}{\nm} (\qty{193}{\THz}), these integrated optical components are becoming increasingly mature for operation at shorter wavelengths to reach the visible spectrum~\cite{sorace-agaskar_versatile_2019,blumenthal_photonic_2020} and interface with other atomic, molecular, and optical systems~\cite{RoppLightSciAppl2023}. Thanks to improvements in propagation loss, high quality factor resonators, in particular microring~\cite{LuNat.Phys.2019, SinclairOpt.ExpressOE2020a} and photonic crystal~\cite{MartinJ.Opt.2017} geometries, now allow for low-power on-chip nonlinear operation for generation of new optical frequencies~\cite{MartyNat.Photonics2021, LuOptica2019, StoneNat.Photon.2023}, including wavelengths reaching the visible. In particular, the prospect of frequency comb generation in the visible spectrum could find direct application in compact optical clockworks for atomic references~\cite{NewmanOptica2019, YuPhys.Rev.Applied2019a,MoilleNature2023} and on-chip spectroscopy systems~\cite{Long2023}. Regardless of the applications and whether they function in linear or nonlinear regimes, the material used for integrated photonics must be transparent at the wavelength of interest, with a higher refractive index than the surrounding material, ensuring light confinement. Silicon nitride (\ce{Si3N4}) is a material of choice~\cite{sorace-agaskar_versatile_2019,sacher_visible-light_2019}, as its refractive index exceeds that of silicon dioxide  (\ce{SiO2}) allowing for total internal reflection, and hence mode-guiding, for a \ce{SiO2}-embeded \ce{Si3N4} photonic structure, with several different fabrication approaches providing large microring resonator quality factor ($Q$)~\cite{PfeifferOpticaOPTICA2018,ji_methods_2021}. In addition, its transparency window reaches far into the visible, and the platform is being developed for mass-scale production capacity~\cite{LiuNatCommun2021a}. Its nonlinear coefficient is high enough to allow $<$\qty{1}{\mW} nonlinear effects~\cite{ji_methods_2021,LuOptica2019,ChangNat.Commun.2020} to be realized in high-$Q$ resonators. However, high-$Q$ and large effective nonlinearity are not sufficient, as parametric nonlinear processes must satisfy energy and momentum conservation, which for bright, broadband frequency comb generation usually necessitates the pump laser frequency be within the \textit{anomalous} dispersion regime. In the context of microrings, the resonator dimensions, specifically their width (RW) and thickness (H), are typically adjusted to induce a geometric dispersion that offsets the material's intrinsic dispersion, thus achieving the desired anomalous regime. However, designing a microring resonator with anomalous dispersion proves increasingly difficult as the wavelength reduces, especially below \qty{900}{\nm} where \ce{Si3N4} and \ce{SiO2} exhibit predominantly normal dispersion. 

In this work, we demonstrate that an air-top-clad \ce{Si3N4} design presents a unique advantage: it provides the fundamental transverse electric (TE) and magnetic (TM) modes with anomalous dispersion at wavelengths deep into the otherwise unreachable visible spectrum when using a \ce{SiO2}-embedded \ce{Si3N4} platform. Through this simple material stack change, we show that the fundamental TE mode can produce anomalous dispersion down to the green at \qty{515}{\THz}, compared with a maximum of about \qty{310}{\THz} for \ce{SiO2}-embedded \ce{Si3N4} microring resonators. Moreover, despite the relatively low-quality factor we observe in our resonators ($Q\approx 100\times10^{3}$), the increased mode confinement, owing to a greater refractive index contrast in this air-top-clad structure, substantially boosts the effective nonlinear coefficient. This results in microcomb generation over a \qty{125}{\THz} spectrum with only about \qty{150}{\mW} power in the input fiber, with an oxide-cladded facet used for low insertion losses~\cite{MoilleAppl.Phys.Lett.2021a}. Finally, we show that air-clad \ce{Si3N4} microcombs, with an appropriate wavelength pump, could reach down to the blue region of the visible spectrum with a standard rectangular cross section microring resonator.

\begin{figure*}[!t]
 \begin{center}
 \includegraphics[width = \figsize]{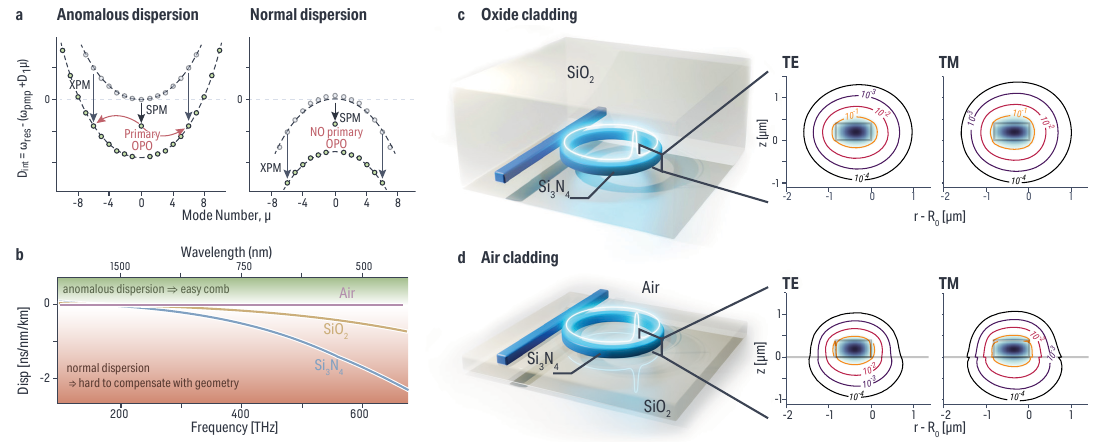}
 \end{center}
 \caption{\label{fig:1}
 \textbf{a} Comparison between anomalous and normal dispersion, which translates into concave and convex integrated dispersion $D_\mathrm{int}$, respectively. The self-phase and cross-phase modulation have a two-fold discrepancy in amplitude, resulting in a different frequency shift between the pumped and remaining modes. In the anomalous case, frequency matching can be fulfilled, allowing for a primary OPO cascading into a comb. In contrast, the normal case precludes such a nonlinear state, preventing the soft excitation of frequency combs in this regime. Thus, anomalous dispersion is necessary for the soft excitation of a bright dissipative Kerr soliton frequency comb. %
 \textbf{b} Material chromatic dispersion. Silicon nitride presents anomalous dispersion in the mid-IR before quickly shifting to normal dispersion at frequencies above \qty{190}{\THz}. However, soft-excited microcombs can still be generated because of the role of geometric dispersion. %
 \textbf{c-d} Mode profile for the first order transverse-electric and transverse-magnetic modes of a \ce{SiO2} embedded (c) and air-clad (d) \ce{Si3N4} microring, taken for a ring radius of 23 $\mu$m and at a wavelength of \qty{1060}{\nm}. Though the modes are largely confined inside the ring core in both oxide- and air-clad geometries, the difference between having a normally dispersive cladding (oxide) and a dispersionless cladding (air) is significant enough to enable anomalous dispersion for air-clad devices in spectral regions where oxide-clad geometries remain normal.
 } 
\end{figure*}

Before reviewing the findings, we should discuss the microring resonators' fundamental physics and dispersion engineering. First, it is important to note that the scope of this work is to generate \textit{bright} dissipative Kerr soliton (DKS) microcombs, which contrary to their dark counterparts, enables broadband comb generation spanning up to or above an octave~\cite{YuPhys.Rev.Applied2019a,MoilleNat.Commun.2021a,liu2021aluminum,MoilleNature2023}. A primary optical-parametric-oscillation (OPO) state must first be reached to \textit{soft-excite} a bright dissipative Kerr soliton~\cite{HanssonPhys.Rev.A2013b}. A red wavelength shift, or a negative frequency shift, is caused by the sign of the Kerr nonlinearity in \ce{Si3N4}. In addition, a two-fold disparity in the nonlinear frequency shift between the pumped mode and its neighbors is caused by the difference between self and cross-phase modulation (SPM and XPM, respectively). Thus, for the primary OPO to exist, the system's dispersion must be anomalous to ensure that energy conservation is satisfied~[\cref{fig:1}(a)].
In most cases, it is more practical to describe the system with its integrated dispersion $D_\mathrm{int}(\mu) = \omega_\mathrm{res}(\mu) - (\omega_\mathrm{pmp} + D_1 \mu) = \sum_{k\ge2} \frac{D_k}{k!}\mu^k$, with $\mu$ the mode number referenced to the pumped one, $\omega_\mathrm{res} (\mu)$ the linear resonant frequencies, $D_1$ the repetition rate at the pump, and $D_k$ the higher order dispersion terms. The dispersion regime is defined by the integrated dispersion curvature $\partial^2_\mu D_\mathrm{int}(\mu)\equiv D_2$, which is normal when convex (\textit{i.e.} $D_2<0$) and anomalous when concave (\textit{i.e.} $D_2>0$). Accounting for SPM and XPM with the integrated dispersion, it is self-evident to notice the need for anomalous dispersion to yield a primary OPO condition. In contrast, normal dispersion does not allow for such conditions, thus making it impossible to soft-excite the system [\cref{fig:1}(a)]. The latter, however, can generate a dark soliton under hard excitation (avoided mode crossing, pulse pumping, etc.)~\cite{liu2014investigation, xu2021frequency, YuNat.Photonics2021}. However, the demonstration of broadband combs in this regime (e.g., towards octave-spanning) still needs to be proven, and the physics of dark pulses is outside the purview of this manuscript. 

A microring resonator's dispersion comprises two contributions: the material dispersion resulting from the intrinsic chromatic refractive index dispersion and the geometric dispersion arising from the variation in light confinement. The first is directly tied to the material's transparency window, with glass (SiO$_2$) and Si$_3$N$_4$ layers exhibiting low-loss transparency from the mid-infrared to the visible before becoming opaque at shorter (ultraviolet) wavelengths [\cref{fig:1}(b)].  As a result, the dispersion of these materials becomes increasingly normal the shorter the wavelength, particularly below \qty{900}{\nm}. However, in a \ce{Si3N4} microring fully encased in \ce{SiO2}, where material dispersion is already normal, bright octave-spanning DKS microcombs pumped at \qty{1060}{\nm} have been observed~\cite{YuPhys.Rev.Applied2019a}. This results from the geometric dispersion, that is, variation of the effective refractive index -- which can be significantly different from the core guiding material -- as a function of wavelength due to modal confinement, and which can compensate for the material dispersion. The effective refractive index can be understood in a qualitative, approximate fashion as related to the optical mode's overlap integral over the material permittivity of the structure. Although the mode profiles of oxide-clad and air-clad structures (\textit{i.e.}, where the \ce{Si3N4} core is sitting on \ce{SiO2} but not embedded in it) are relatively comparable [\cref{fig:1}(c)-(d)], their effective refractive indices at the same wavelength will be significantly different due to the mode in the air-clad structure being more confined in the guiding material. The geometric dispersion of the \ce{SiO2}-encapsulated \ce{Si3N4} microring [\cref{fig:1}(c)] needs to be much stronger to compensate for the material dispersion than its top-air clad counterpart [\cref{fig:1}(d)], given that \ce{SiO2} becomes more normal the shorter the wavelength while air is non-dispersive.
  
\begin{figure}[t]
 \begin{center}
 \includegraphics{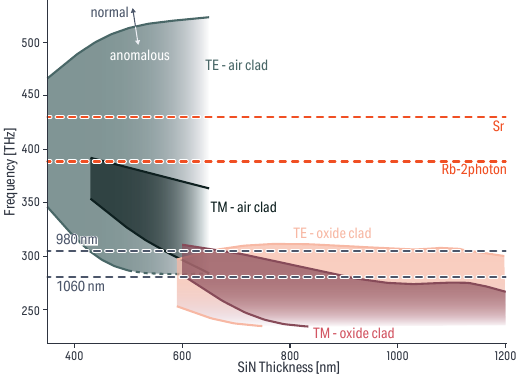}
 \caption{\label{fig:2}
    Limit of anomalous dispersion in the case of fundamental transverse electric (TE) and magnetic (TM) modes for both air and oxide top-cladding. The limit is found for a different set of ring width and thickness parameters, with the limit determined as where the maximum of the dispersion remains positive. The atomic transitions of interest for integrated optical clocks are highlighted in orange, while the pump frequency for current short-wavelength octave spanning comb generation at \qty{1060}{\nm}~\cite{YuPhys.Rev.Applied2019a,moille2022towards} and \qty{980}{\nm}~\cite{moille2022towards} represented with the dark dashed lines. 
 } 
 \end{center}
\end{figure}

\begin{figure*}[!t]
    \begin{center}
    \includegraphics[width = \figsize]{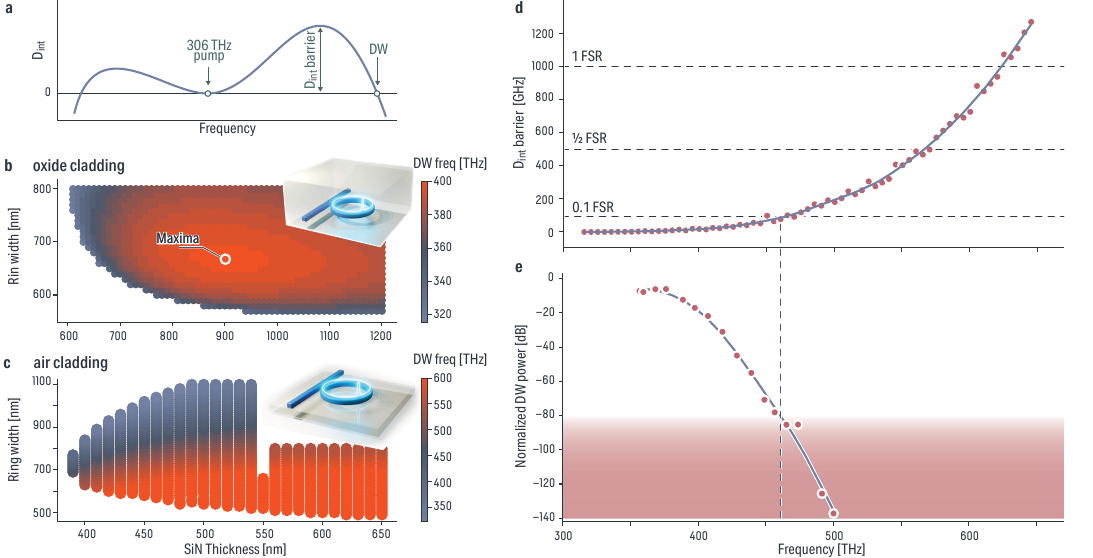}
    \end{center}
    \caption{\label{fig:3}
    \textbf{a} Integrated dispersion ($D_\mathrm{int}$) schematic. The zero crossings represent phase-matched comb teeth with the resonator dispersion, and hence allow for resonant enhancement and the creation of dispersive waves (DWs). Yet, these DWs exist because of the roll-off of $D_\mathrm{int}$ due to higher-order dispersion, which also forms a barrier which, analogous to a tunneling through a potential well, can be used as a metric related to the DW efficiency. %
    \textbf{b-c} DW frequency (color scale) from the zero-crossing of $D_\mathrm{int}$ for a resonator pumped at $306$~THz (\qty{980}{\nm}) for air-clad (b) and oxide-clad (c) \ce{Si3N4} resonators. The ring width and thickness are varied to find the maximum DW frequency, which for the oxide-clad device exhibits a global maximum at about \qty{400}{\THz}. In contrast, the air-clad device does not have a maximum phase-matched frequency within the range studied (up to \qty{600}{\THz}). %
    \textbf{d} Minimum $D_\mathrm{int}$ barrier for a given zero crossing of $D_\mathrm{int}$ for the air-clad structure. The barrier minimum is found by selecting the appropriate $RW$/$H$ combination leading to the DW of interest. The lower the barrier, the easier the DW creation. The free spectral range $D_1$ and its relevant fraction are highlighted in dashed lines. %
    \textbf{e} LLE simulations of the DW power for an air-clad device using the $D_\mathrm{int}$ profile leading to the results in d. We consider 80~dB below the comb maximum as the minimal required power for a DW to be used in a system (either due to SNR or detection), with the red shaded region indicating DW powers that are below this limit. Interestingly, this limit of 80~dB is reached for a  $D_\mathrm{int}$-barrier of about $D_1/10$, highlighting that despite the possibility of achieving a much higher phase-matching frequency, the limit for efficient DW generation with a $306$~THz pump is around \qty{475}{\THz} (vertical dashed line).
    }
   \end{figure*}

\begin{figure}[!t]
    \begin{center}
    \includegraphics{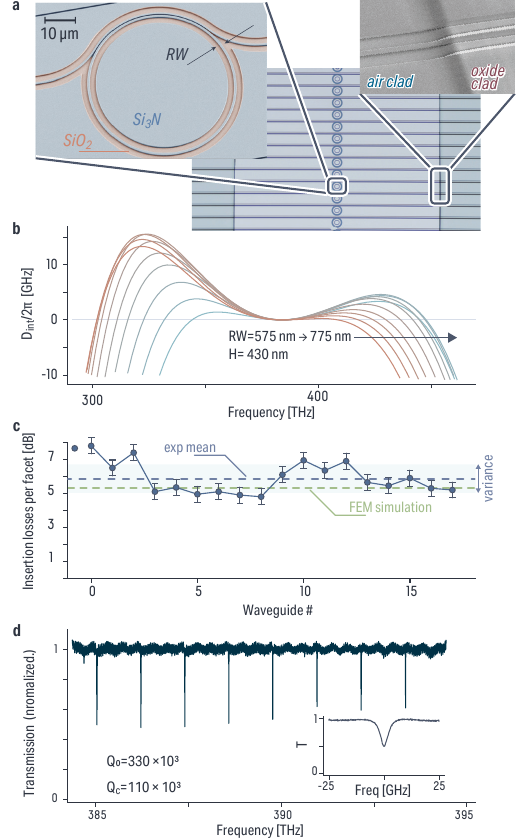}
    \end{center}
    \caption{\label{fig:4}
    \textbf{a} Image of the fabricated microring resonator with air cladding. The microring (left inset; scale bar is 10 $\mu$m) is coupled to a pulley-like waveguide to allow for efficient coupling and extraction of the comb. We use a low temperature PECVD liftoff technique to still provide oxide-clad facets, which enable the tapering down of the access waveguides for efficient input and output coupling to the chip. %
    \textbf{b} Integrated dispersion of the fundamental TM mode for different ring width and a thickness of $H=430$~nm. The dispersion remains anomalous for for the ring width of interest between $RW=575$~nm and $RW=775$~nm, while a significant shift of the high frequency DW of about $10$~THz occurs for ring width shift of about the fabrication tolerance of $25$~nm. %
    \textbf{c} Insertion loss per facet of eighteen devices (blue dots with \qty{\pm 0.5}{\dB} error bars) with a \qty{100}{\nm} width \ce{SiO2}-clad inverse taper, with their mean value (dashed blue line) and the corresponding variance (shaded blue area) compared to the finite element method (FEM) simulated value (green dashed line). The \ce{SiO2} lift-off~\cite{MoilleAppl.Phys.Lett.2021a} enables low insertion losses while supporting air-clad microrings that exhibit anomalous dispersion close to the visible. 
    \textbf{d} Linear characterization of a fabricated microring resonator with \qty{H=430}{\nm} and \qty{RW=650}{\nm} for the fundamental TM mode. A zoom of around a given resonance is displayed as inset with a intrinsic and coupled quality factor of $Q_\mathrm{0}=330\times10^3$ and $Q_\mathrm{c}=110\times10^3$ respectively. %
    }    
\end{figure}
To effectively compare the two structures, we perform dispersion simulations for a wide range of \ce{Si3N4} thickness, ring width and optical frequency, assuming a 23~$\mu$m ring radius that corresponds to a resonator free-spectral range of \qty{1}{\THz}, which is consistent with previous octave-spanning comb demonstratons~\cite{LiOptica2017a,YuPhys.Rev.Applied2019a,MoilleNature2023}. For each geometry, we calculate the dispersion $D=-\frac{\partial^2}{\partial \omega^2}\left(\omega n_\mathrm{eff}\right)$. We then find the peak of the dispersion, and its limit to be positive, hence determining if anomalous dispersion exists for that given geometry. This allows for direct comparison between air and oxide-top cladding structures, and fundamental TE and TM polarization [\cref{fig:2}]. Air cladding shows a significant larger allowed anomalous dispersion frequency, with a maximum \qty{>500}{\THz} for the TE mode and \qty{\approx 390}{\THz} for the TM mode. In contrast, oxide cladding only allows, regardless of mode polarization, a frequency of \qty{\approx 315}{\THz} for the maximum of the dispersion to still be anomalous. We note that these anomalous frequency limits are both ring width and thickness dependent and only the optimum ring width is utilized in~\cref{fig:2}.

The anomalous regime is critically important to achieve bright DKS generation at the pump frequency. However, higher-order dispersion allows one or several phase-matching frequencies, where the comb teeth are resonantly enhanced~\cite{BraschScience2016}. Per definition, these dispersive waves (DWs) will exist in the normal dispersion regime of the resonator and could significantly increase the bandwidth of the comb~\cite{LiOptica2017a, PfeifferOpticaOPTICA2017}. Thus, it is also critical to study the possibility to achieve DW operation at the highest frequency possible. Following the previous simulation protocol, we vary both ring width and thickness of the microring resonator for air and oxide-top cladding while computing the integrated dispersion $D_\mathrm{int}$ with $\omega_\mathrm{pmp}\approx 306$~THz ($\approx 980$~nm) the chosen pump frequency [\cref{fig:3}a]. When $D_\mathrm{int} = 0$, the frequency of resonance is aligned with the fixed frequency grid defined by  $(\omega_\mathrm{pmp} + \mu D_1 )$, hence the DKS comb; thus, the DW is generated when the comb tooth is on resonance. %
In the case of oxide-top cladding with a $23$~{\textmu}m ring radius (about \qty{1}{\THz} free spectral range), a global maximum appears for a DW at a frequency of about \qty{400}{\THz} for $RW=900$~nm and $H=670$~nm [\cref{fig:3}b], highlighting the limited achievable high frequency comb teeth in this system. In contrast, the air cladding does not present a maximum in the simulation range studied, with phase-matching creating potential DWs up to \qty{600}{\THz} (i.e. \qty{500}{\nm}) [\cref{fig:3}c]. 

\begin{figure*}[t]
    \begin{center}
    \includegraphics[width = \figsize]{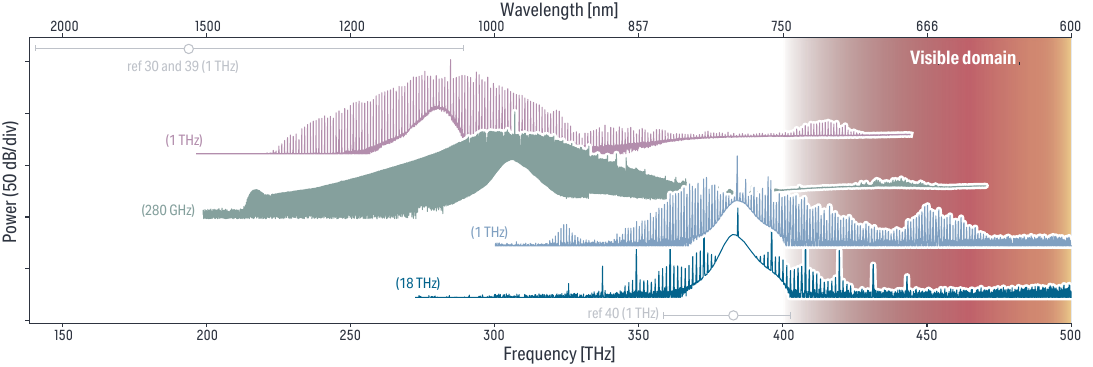}
    \end{center}
    \caption{\label{fig:5}
    Experimental demonstrations of microcomb generation in the visible using air-clad resonators. The top spectrum (purple) is of a \qty{1}{\THz} free spectral range resonator that is pumped on a fundamental TE mode at \qty{283}{\THz}, resulting in a broadband comb whose dispersive wave allows access to visible wavelengths at \qty{420}{\THz} (\qty{714}{\nm}). Shorter wavelengths can be accessed by shifting the pump to \qty{305}{\THz}, where an octave-spanning microcomb at an FSR of \qty{280}{\GHz} is generated (green), with a total comb span of \qty{220}{\THz} and a high frequency DW at \qty{440}{\THz} (\qty{681}{\nm}). To push further into the visible, the fundamental TM mode of a dispersion-engineered resonator can be pumped at \qty{384}{\THz} (\qty{780}{\nm}), generating a dual-DW microcomb (light blue) reaching \qty{320}{\THz} (\qty{937}{\nm}) and \qty{450}{\THz} (\qty{666}{\nm}), with a total bandwidth of \qty{130}{\THz} (light blue). Beyond these modulation instability combs, other type of microcomb states, including likely soliton crystal states (bottom dark blue spectrum) can be generated, hinting that single DKS states could be generated under the right conditions.  %
    } 
\end{figure*}

\begin{figure}[!b]
    \begin{center}
    \includegraphics[width = 0.95\columnwidth]{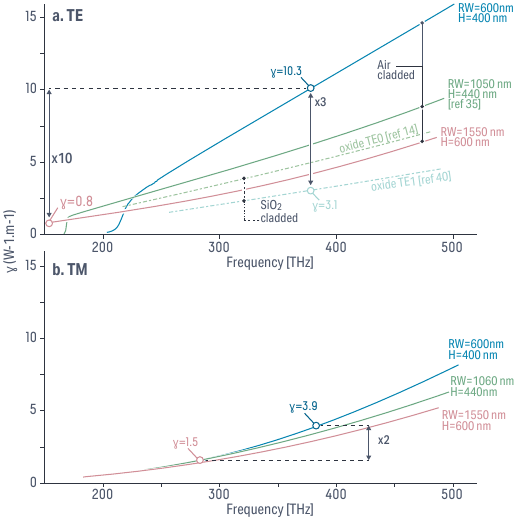}
    \end{center}
    \caption{\label{fig:6}
        Dependence of the effective nonlinear coefficient $\gamma$ on frequency and geometry. The linear dependence of $\gamma$ on $\omega$ leads to a natural increase for higher frequency (shorter wavelength) pump microcomb operation. Dispersion engineering to allow for anomalous dispersion for visible-pumped microcombs makes the microring ring cross-section smaller than they may be otherwise, thereby increasing the modal confinement and hence reducing $A_\mathrm{eff}$. The combination of both effects allows for a significant increase in $\gamma$ for visible pumped microcombs, which can compensate for lower quality factor. %
        \textbf{a} and \textbf{b} are for fundamental TE and TM modes, respectively, for a variety of air-clad (solid lines) and SiO$_2$ clad (dashed line) geometries. %
    } 
\end{figure}

However, the presence of a phase matching frequency due to higher-order dispersion does not guarantee the creation of an efficient DW, since the comb power available at this frequency needs to be sufficient for appreciable resonance enhancement. A simple metric to define if a DW can be efficiently created is to study the integrated dispersion barrier through which the soliton must traverse to eject radiation at the DW. Similar to a potential well, the lower the barrier the easier the tunneling effect. We report in \cref{fig:3}d the \textit{minimum} integrated dispersion barrier for a given DW frequency for the air-clad device. The larger the DW frequency reaches, the higher the barrier becomes and can become commensurate with the FSR. Using the Lugiato-Lefever Equation (LLE)~\cite{ChemboPhys.Rev.A2013} model through the \textit{pyLLE} freeware~\cite{MoilleJ.RES.NATL.INST.STAN.2019}, we compute the DW power against the comb maximum [\cref{fig:3}e]. We assume that a DW that is more than 80~dB below the comb maximum would not be experimentally  interesting since its power would be too low to detect (or to be useful). Interestingly, we notice that this limit appears for a $D_\mathrm{int}$ barrier of $\approx~D_1/10$, which results in a maximum DW frequency of about \qty{475}{\THz} (i.e. \qty{630}{\nm}) for the air-clad device. This is a significant improvement compared to the oxide-clad device, and also highlights the importance of not only finding phase-matching frequencies but also the efficiency of the DW generated through the $D_\mathrm{int}$ barrier.  

We proceed to try and experimentally demonstrate that the top air-clad device can reach a shorter wavelength than that predicted for its oxide-embedded counterpart. We use microring resonators with anomalous dispersion with the fundamental TE mode at \qty{284}{\THz} (\qty{1060}{\nm}), at \qty{306}{\THz} (\qty{980}{\nm}) following reference~\cite{moille2022towards}, and at \qty{380}{\THz} (\qty{780}{\nm}) in the fundamental TM mode. Each of these resonators is made using stoichiometric silicon nitride grown with a 7:1 ammonia-dichlorosilane gas ratio~\cite{MoilleOpt.Lett.OL2021a}. To mitigate insertion losses to and from the chip with lensed optical fibers, we use an inverse-tapered waveguide design with oxide-top cladding, while the resonator is still air-clad thanks to a liftoff process enabled by low temperature plasma-enhanced chemical deposition (PECVD) of the \ce{SiO2}  [\cref{fig:4}a]. To accommodate fabrication constraints, we select a \qty{100}{\nm} inverse-taper width instead of the optimal \qty{50}{\nm}. This choice increases expected insertion losses from \qty{2}{\dB} to \qty{5}{\dB} per facet. Lithography and etching variations may explain why measured performance slightly exceeds finite-element method simulations.\\
The system's fundamental TM mode dispersion presents a strong dependence on the structure geometry, particularly the ring width, as expected by the shorter operation wavelength. Assuming a pump mode at $\omega_\mathrm{pmp}=384\times2\pi$~THz and considering a ring width range between $RW=575$~nm and $RW=775$~nm, a change in $RW$ of about \qty{25}{\nm} can produce a shift of the high-frequency DW up to \qty{10}{\THz} [\cref{fig:4}b]. Given this sensitivity, we are not attempting to compare our experimental results of the produced DW directly; rather, we are using these results as a guideline to obtain DWs in the relevant spectral region, which is anticipated to be between \qty{420}{\THz} and \qty{450}{\THz}. Upon fabrication, using electron beam lithography to define the structure followed by standard reactive ion etching using \ce{CHF3}, and the low-temperature PECVD SiO$_2$ lift-off, we obtained insertion loss per facet of about 5~dB for the fundamental TM modes and an inverse taper width of \qty{120}{\nm} most for the 18 devices tested [\cref{fig:4}(c)].  This matches well with the expected value obtained through mode-overlap finite element method (FEM) simulations with a 2.5~{\textmu}m waist diameter lensed fiber mode. The spectroscopy of the microring resonators exhibits a coupled and intrinsic factors of $Q_\mathrm{c}=110\times10^3$ and $Q_0=330\times10^3$ respectively, at \qty{780}{\nm}~[\cref{fig:4}(d)]. Although these values are lower by about a factor of 5 compared to state-of-the-art $Q$ at \qty{780}{\nm} of \ce{Si3N4} embedded in \ce{SiO2}~\cite{SinclairOpt.ExpressOE2020} or with air cladding~\cite{LuOpticaOPTICA2020}. This discrepancy likely stems from the small thickness and ring width, which tightly confine the optical mode and amplify the effects of fabrication-induced sidewall roughness losses. Nevertheless, our devices maintain $Q$ factors sufficient for nonlinear optics, particularly given the enhanced effective nonlinearity arising from high frequency and small cross-section modal area, as discussed later. %
From spectroscopic measurements of the microring, only $D_2$ is accessible, providing qualitative information about normal or anomalous dispersion. Higher-order dispersion terms---critically important to create broadband combs through dispersive waves---remain inaccessible due to the need for laser wavelength spans beyond that which can be accessed with commercially available continuously tunable lasers. However, the comb spectrum itself, through the phase-matching condition leading to DWs, offers valuable information.  

\begin{figure*}[t]
    \begin{center}
    \includegraphics[width = \figsize]{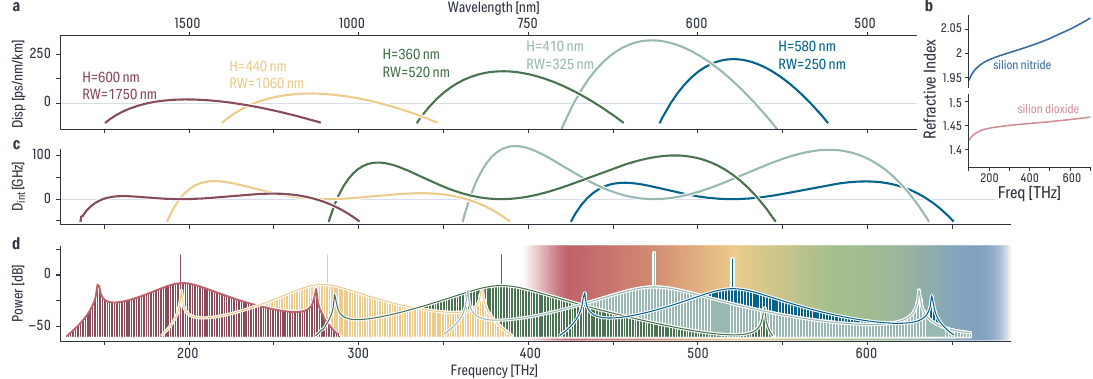}
    \end{center}
    \caption{\label{fig:7}
Theoretical microcomb generation, obtained from the dispersion simulations and the LLE, available with air-clad \ce{Si3N4} devices. Given the flexible dispersion engineering that allows for anomalous dispersion from C-Band operation down to deep in the visible, broad bandwidth combs spanning up to or close to an octave are available with dual-DW microcomb generation across a broad range of frequencies. %
\textbf{a} Dispersion simulations using the finite element method accounting for accurate material dispersion as the solver uses \textbf{b} experimentally obtained material refractive indices for silicon nitride (blue) and silicon dioxide (pink). %
\textbf{c} Integrated dispersion $D_\mathrm{int}$, highlighting the two zero-crossings for dual DW generation. The pump frequencies and resonator geometries have been chosen to achieve roughly symmetric $D_\mathrm{int}$ profiles. %
\textbf{d} LLE simulated microcomb spectrum for each dispersion-engineered \qty{1}{\THz} FSR microring resonator demonstrating the silicon nitride platform capability to generate visible microcombs down to blue wavelengths thanks to the absence of dielectric top cladding. %
    } 
\end{figure*}

We next characterize the nonlinear performance of the above air-clad devices (Fig.~\ref{fig:5}). We focus on modulation instability (MI) combs as our focus is on the shortest accessible wavelength.  First, using the fundamental TE mode of a microring pumped with \qty{\approx150}{\mW} on-chip power at \qty{283}{\THz}, with a ring radius \qty{RR = 23}{\um} ring width \qty{RW = 980}{\nm}, and ring thickness \qty{H = 430}{\nm}, comb tooth frequencies as high as \qty{420}{\THz} (\qty{714}{\nm}) are reachable thanks to a dispersive wave generation while pumping the microring at \qty{283}{\THz} (\qty{1060}{\nm}). Although this comb's short wavelength reach is substantial, the absence of a low frequency DW prevents it to be further used for octave-spanning operation, despite the overall large bandwidth. It is important to point out the difference in spectral bandwidth between other broadband microcombs. In particular, an octave span will be much smaller for a microcomb pumped in the C-band around \qty{193}{\THz}, where harmonic DWs could be at \qty{140}{\THz} and \qty{280}{\THz} (similar to ref~\cite{LiOptica2017a}), resulting in a total span of \qty{140}{\THz}. In the current case, by reaching to the visible domain which is at a much higher frequency, the total required span reach an octave needs to be doubled, which is a significant challenge. We succeed in this task by pumping a larger ring resonator with \qty{RR = 82}{\um} (corresponding to an FSR of $\approx$~\qty{280}{\GHz}) and $\{RW=875\,\mathrm{nm}; H=410\,\mathrm{nm}\}$~nm at a pump frequency around \qty{305}{\THz} and larger on-chip pump power of \qty{400}{\mW} because of the larger modal volume compared to the previous device under test, enabling both a short DW at \qty{440}{\THz} and a long DW at \qty{220}{\THz}. This corresponds to an octave span with a total bandwidth of up to \qty{220}{\THz}. Such large bandwidth is made possible in part due to the \qty{L_\mathrm{c}\approx{12}}{\um} pulley-like coupler that enables a relatively flat coupling dispersion and efficient short wavelength extraction~\cite{MoilleOpt.Lett.2019b}.
As presented earlier in~\cref{fig:3}, reaching even shorter wavelengths while maintaining a pump in the \qty{300}{\THz} band is challenging since the DW power will considerably decrease. To reach shorter wavelengths, we continue to take advantage of the air-clad resonator geometry, with  a similar system as shown in~\cref{fig:4} but pumped at a higher frequency and using the fundamental TM mode. We pumped our $\{RW=625\,\mathrm{nm}; H=430\,\mathrm{nm}\}$ air-clad \ce{Si3N4} microring resonator at \qty{384}{\THz} with about \qty{150}{\mW} of power on-chip, a value comparable with that used in octave-spanning combs~\cite{LiOptica2017a, SpencerNature2018}, as the lower quality factor is balanced out by the larger effective nonlinearity. The obtained modulation instability frequency comb presents two DWs at about \qty{320}{\THz} and \qty{450}{\THz}. The short DW at \qty{450}{\THz} (\textit{i.e.} \qty{666}{\nm}), well within the visible spectrum, presents a considerable power thanks to the $L_\mathrm{15}$~{\textmu}m pulley-like coupler. Compared to the existing state-of-the-art \ce{Si3N4} microcombs, this represents the shortest wavelength achieved, about \qty{50}{\THz} higher than ref~\cite{ZhaoOpticaOPTICA2020} (pumped at \qty{780}{\nm}) and \qty{20}{\THz} higher than ref.~\cite{moille2022towards} (pumped at \qty{1060}{\nm}). Although our microcomb is far from octave-spanning, its span $>$\qty{130}{\THz} is comparable to that of telecom-band octave-spanning combs, and its bandwidth is multiple times larger than previously demonstrated \qty{780}{\nm} pumped \ce{Si3N4} combs. The latter is due to the use of  a fundamental mode, which simultaneously presents shallower dispersion and higher nonlinearity than higher-order mode schemes. Although we could not produce a DKS state in this work, primarily due to the large thermo-refractive bistability that limits soliton access and which we believe is due to residual linear absorption from the silicon nitride, we were able to access a state that from its comb spectrum may be a soliton crystal (bottom dark blue trace in Fig.~\ref{fig:5}), whose thermal stability requirements are much less stringent than that of a DKS due to their comparatively high intracavity power. Further noise and coherence measurements must be carried out to fully characterize this multi-DKS state.

The effective nonlinearity $\gamma = \frac{n_\mathrm{2}\omega}{A_\mathrm{eff}c}$, with nonlinear refractive index $n_2 = 2.5\times 10^{-15}$~W$\cdot$cm\textsuperscript{-2}, $n_\mathrm{eff}$ the effective refractive index, and  $A_\mathrm{eff}$ the mode effective area, is at the same time frequency and confinement dependent [\cref{fig:6}]. Recalling that this coefficient plays a crucial role in the LLE for the magnitude of the self-phase modulation term $i\gamma L|a|^2 a$, the higher the $\gamma$ value the lower the pump power needed to trigger a nonlinear state. The linear dependence of $\gamma$ on $\omega$ automatically increases its value at high frequency, making nonlinear operation and comb generation more efficient at short wavelength. In addition to the direct proportionality between $n_2$ and $\omega$, the effective area $A_\mathrm{eff}$ is also frequency, geometry, and material dependent, making the relationship between $\gamma$ and the system under study non-trivial. From the mode profile first shown in~\cref{fig:1}c-d, it is evident that air-clad structures exhibit a tighter mode confinement than their oxde-clad counterparts, since the refractive index contrast between the core and cladding is larger. To study this effect, we solve the cross-section modes of different ring resonator geometries and material stacks to determine the behavior of $\gamma$. We use air-clad structures of $\{RW=1550; H=600\}$~nm, $\{RW=1050; H=440\}$~nm and $\{RW=600, H=400\}$~nm, corresponding to octave-spanning combs pumped at \qty{1550}{\nm} and producing a DW at \qty{1060}{\nm}~\cite{LiOptica2017a}, octave-spanning combs pumped at \qty{1060}{\nm} with a DW reaching \qty{780}{\nm}~\cite{MoilleOpt.Lett.OL2021a}, and our current dispersion design, respectively. We use \ce{SiO2}-embedded microrings corresponding to an octave-spanning comb pumped at 1060nm and reaching a DW at \qty{780}{\nm}~\cite{YuPhys.Rev.Applied2019a}, and a second-order TE mode pumped at \qty{780}{\nm} and reaching down to \qty{720}{\nm} without a DW~\cite{ZhaoOpticaOPTICA2020}. The smaller size of the ring cross-section for the visible dispersion-engineered microring allows for tighter confinement, reducing the effective area $A_\mathrm{eff}$ and allowing for about a factor of 10$\times$ increase in $\gamma$ for a fundamental TE mode between an air-clad structure pumped at \qty{780}{\nm} for broadband comb generation and a reference structure pumped at \qty{1550}{\nm}. A factor 3$\times$ is seen between a \qty{1060}{\nm} pumped and \qty{780}{\nm} pumped comb, showcasing an increase larger than that expected from the scaling of $\gamma$ on $\omega$ alone. For the purpose of this work, it is important to emphasize that the choice of an air-clad resonator in comparison to a \ce{SiO2}-embedded structure not only allows for shorter wavelength anomalous dispersion, but also allows for a much tighter $A_\mathrm{eff}$, resulting in an increase in $\gamma$. Finally, although a short wavelength pumped \ce{SiO2} microcomb has been demonstrated~\cite{ZhaoOpticaOPTICA2020}, it is important to point out that $\gamma$ also depends on the mode considered. Indeed, a higher-order mode will be less confined than a fundamental mode, resulting in a reduced $A_\mathrm{eff}$. Therefore, fundamental pumping for microcomb applications will showcase an advantage in the nonlinear interaction strength. For example, in \cref{fig:6}(a), we show how the fundamental TE mode has $\gamma$ that is more than 3$\times$ larger than that of the first higher-order mode (at \qty{780}{\nm}), while the fundamental TM mode $\gamma$ is also slightly larger. This increase in $\gamma$ by using fundamental mode, air-clad resonators is largely responsible for the demonstrated broadband, double DW comb generation at \qty{150}{\mW} of on-chip power, despite the close to one order-of-magnitude smaller total quality factor compared to C-band microcombs.  The total quality factors are also $\approx5\times$ smaller than the state-of-the-art at \qty{780}{\nm} for thick air-clad \ce{Si3N4} microrings~\cite{LuOpticaOPTICA2020} and \ce{SiO2}-embedded thick \ce{Si3N4} microrings~\cite{SinclairOpt.ExpressOE2020}.

Using finite element method dispersion modeling alongside the LLE, we can find the limits of the achievable comb span and frequency that can be reached by pumping integrated air-clad microring resonators at different frequencies. Using the dispersion mapping presented in~\cref{fig:2}, we retrieve the maximum pumped frequency achievable for a symmetric $D_\mathrm{int}(\mu)$. It is important to note that for such a high frequency, the consideration of realizing octave-spanning bandwidths without some further advance seems unrealistic, given that it would involve a microcomb span of more than \qty{300}{\THz}, which to this day has yet to be demonstrated. We proceed and calculate the different theoretical DKS spectra for different pumps, including \qty{193}{\THz}~\cite{LiOptica2017a}, \qty{283}{\THz} (from ref~\cite{moille2022towards}), \qty{383}{\THz} (optimal fundamental TE mode presented here), \qty{475}{\THz}, and \qty{517}{\THz} [\cref{fig:7}]. These calculations highlight the potential for visible \ce{Si3N4} microcombs to access wavelengths deep in the visible from experimentally measured material dispersion, with the full range of designs covering any frequency between \qty{150}{\THz} up to \qty{640}{\THz}, the latter comb being produced thanks to anomalous dispersion for the pump in the green at \qty{515}{\THz}. It is interesting to point out that---despite the lack of octave spans in the design proposed here, in particular since the bandwidth of an octave in the visible becomes very large ---we propose a \ce{Si3N4} pumped at \qty{475}{\THz} which can span over \qty{275}{\THz} (from about \qty{360}{\THz} to \qty{635}{\THz}), making it already twice as broad as state-of-the-art octave-spanning combs pumped at \qty{193}{\THz}. Though there are several challenges to overcome in realizing such performance in practice, including achieving sufficiently high resonator $Q$s, limiting residual linear absorption and improving thermal stability, preventing potential two- or three-photon absorption, and identifying suitable pump laser sources to drive the resonators, recent development of high-$Q$, high-confinement \ce{Si3N4} photonic devices at visible wavelengths~\cite{stone_efficient_2022,Corato-Zanarella:24} suggests a path towards experimental realization.

In conclusion, we have discussed the importance of resonator design for achieving \ce{Si3N4}-based microcombs for visible wavelength applications. To achieve anomalous dispersion at relevant wavelengths -- indispensable for bright dissipative Kerr soliton generation that results in broadband combs -- the resonator cladding material choice is particularly important since \ce{SiO2} presents increasingly strong normal dispersion as the wavelength gets shorter. We show that the geometric dispersion provided by an air-clad resonator geometry is able to bypass the limitations of an \ce{SiO2}-embedded counterpart, so that broadband anomalous dispersion and phase-matched dispersive wave generation in the visible are possible. As noted earlier, air-clad geometries have the further advantages of allowing for post-processing to fine tune dispersion~\cite{MoilleAppl.Phys.Lett.2021a} and can be effectively controlled using underlying integrated micro-heaters~\cite{MoilleAPLPhotonics2022}. Using this air-clad geometry, we theoretically show that symmetric anomalous dispersion up to a pump frequency of \qty{517}{\THz} within the visible spectra is possible. Experimentally, we show modulation instability combs that reach an unprecedented high frequency in the red up to \qty{450}{\THz} (\qty{666}{\nm}). We further highlight that such comb generation is possible despite the relatively high resonator losses we observe, which are compensated by a significant increase in the nonlinear coefficient $\gamma$ in comparison to shorter wavelengths. This allows for comb generation with an on-chip power of about \qty{150}{\mW}, which is similar to that used in previous octave-spanning comb demonstrations at longer wavelengths. Finally, we highlight that the air-clad geometries should support broadband comb generation at frequencies up to \qty{650}{\THz} (\qty{461}{\nm}). Our work suggests that the current state-of-the art in integrated microcombs is yet to reach its full potential when it comes to visible wavelength access. Further improvement of the losses and available pump laser wavelengths will allow for such combs to be realized in practice, enabling a range of applications, including direct interfacing to many atomic transitions.

\vspace{1em}
\noindent \textbf{\large Acknowledgments}\\
\noindent We acknowledge partial funding support from the Space Vehicles Directorate of the Air Force Research Laboratory, the Atomic–Photonic Integration programme of the Defense Advanced Research Projects Agency, and the NIST-on-a-chip program of the National Institute of Standards and Technology. We thank Daniel Hickstein and Wenqi Zhu for insightful feedback. 

\noindent G.M dedicates this work to T.B.M -- always.

\vspace{1em}
\noindent \textbf{\large Conflict of Interest} \\
The authors declare no conflict of interest

\vspace{1em}
\noindent \textbf{\large Data Availability Statement} \\
The data that supports the plots within this paper and other findings of this study are available from the corresponding authors upon request. The simulation code is available from the authors through the pyLLE package available online~\cite{MoilleJ.RES.NATL.INST.STAN.2019}, with a modification that is available upon reasonable request, using the inputs and parameters presented in this work.\\

\bibliographystyle{apsrev4-2}

\begin{thebibliography}{43}%
    \makeatletter
    \providecommand \@ifxundefined [1]{%
     \@ifx{#1\undefined}
    }%
    \providecommand \@ifnum [1]{%
     \ifnum #1\expandafter \@firstoftwo
     \else \expandafter \@secondoftwo
     \fi
    }%
    \providecommand \@ifx [1]{%
     \ifx #1\expandafter \@firstoftwo
     \else \expandafter \@secondoftwo
     \fi
    }%
    \providecommand \natexlab [1]{#1}%
    \providecommand \enquote  [1]{``#1''}%
    \providecommand \bibnamefont  [1]{#1}%
    \providecommand \bibfnamefont [1]{#1}%
    \providecommand \citenamefont [1]{#1}%
    \providecommand \href@noop [0]{\@secondoftwo}%
    \providecommand \href [0]{\begingroup \@sanitize@url \@href}%
    \providecommand \@href[1]{\@@startlink{#1}\@@href}%
    \providecommand \@@href[1]{\endgroup#1\@@endlink}%
    \providecommand \@sanitize@url [0]{\catcode `\\12\catcode `\$12\catcode `\&12\catcode `\#12\catcode `\^12\catcode `\_12\catcode `\%12\relax}%
    \providecommand \@@startlink[1]{}%
    \providecommand \@@endlink[0]{}%
    \providecommand \url  [0]{\begingroup\@sanitize@url \@url }%
    \providecommand \@url [1]{\endgroup\@href {#1}{\urlprefix }}%
    \providecommand \urlprefix  [0]{URL }%
    \providecommand \Eprint [0]{\href }%
    \providecommand \doibase [0]{https://doi.org/}%
    \providecommand \selectlanguage [0]{\@gobble}%
    \providecommand \bibinfo  [0]{\@secondoftwo}%
    \providecommand \bibfield  [0]{\@secondoftwo}%
    \providecommand \translation [1]{[#1]}%
    \providecommand \BibitemOpen [0]{}%
    \providecommand \bibitemStop [0]{}%
    \providecommand \bibitemNoStop [0]{.\EOS\space}%
    \providecommand \EOS [0]{\spacefactor3000\relax}%
    \providecommand \BibitemShut  [1]{\csname bibitem#1\endcsname}%
    \let\auto@bib@innerbib\@empty
    \bibitem [{\citenamefont {Cole}\ \emph {et~al.}(2009)\citenamefont {Cole}, \citenamefont {Huebner},\ and\ \citenamefont {Johnson}}]{ColeIEEECommun.Mag.2009}%
      \BibitemOpen
      \bibfield  {author} {\bibinfo {author} {\bibfnamefont {C.}~\bibnamefont {Cole}}, \bibinfo {author} {\bibfnamefont {B.}~\bibnamefont {Huebner}},\ and\ \bibinfo {author} {\bibfnamefont {J.~E.}\ \bibnamefont {Johnson}},\ }\href {https://doi.org/10.1109/MCOM.2009.4804385} {\bibfield  {journal} {\bibinfo  {journal} {IEEE Communications Magazine}\ }\textbf {\bibinfo {volume} {47}},\ \bibinfo {pages} {S16} (\bibinfo {year} {2009})}\BibitemShut {NoStop}%
    \bibitem [{\citenamefont {Fahrenkopf}\ \emph {et~al.}(2019)\citenamefont {Fahrenkopf}, \citenamefont {McDonough}, \citenamefont {Leake}, \citenamefont {Su}, \citenamefont {Timurdogan},\ and\ \citenamefont {Coolbaugh}}]{FahrenkopfIEEEJ.Sel.Top.QuantumElectron.2019}%
      \BibitemOpen
      \bibfield  {author} {\bibinfo {author} {\bibfnamefont {N.~M.}\ \bibnamefont {Fahrenkopf}}, \bibinfo {author} {\bibfnamefont {C.}~\bibnamefont {McDonough}}, \bibinfo {author} {\bibfnamefont {G.~L.}\ \bibnamefont {Leake}}, \bibinfo {author} {\bibfnamefont {Z.}~\bibnamefont {Su}}, \bibinfo {author} {\bibfnamefont {E.}~\bibnamefont {Timurdogan}},\ and\ \bibinfo {author} {\bibfnamefont {D.~D.}\ \bibnamefont {Coolbaugh}},\ }\href {https://doi.org/10.1109/JSTQE.2019.2935698} {\bibfield  {journal} {\bibinfo  {journal} {IEEE Journal of Selected Topics in Quantum Electronics}\ }\textbf {\bibinfo {volume} {25}},\ \bibinfo {pages} {1} (\bibinfo {year} {2019})}\BibitemShut {NoStop}%
    \bibitem [{\citenamefont {Thomson}\ \emph {et~al.}(2016)\citenamefont {Thomson}, \citenamefont {Zilkie}, \citenamefont {Bowers}, \citenamefont {Komljenovic}, \citenamefont {Reed}, \citenamefont {Vivien}, \citenamefont {{Marris-Morini}}, \citenamefont {Cassan}, \citenamefont {Virot}, \citenamefont {F{\'e}d{\'e}li}, \citenamefont {Hartmann}, \citenamefont {Schmid}, \citenamefont {Xu}, \citenamefont {Boeuf}, \citenamefont {O'Brien}, \citenamefont {Mashanovich},\ and\ \citenamefont {Nedeljkovic}}]{ThomsonJ.Opt.2016a}%
      \BibitemOpen
      \bibfield  {author} {\bibinfo {author} {\bibfnamefont {D.}~\bibnamefont {Thomson}}, \bibinfo {author} {\bibfnamefont {A.}~\bibnamefont {Zilkie}}, \bibinfo {author} {\bibfnamefont {J.~E.}\ \bibnamefont {Bowers}}, \bibinfo {author} {\bibfnamefont {T.}~\bibnamefont {Komljenovic}}, \bibinfo {author} {\bibfnamefont {G.~T.}\ \bibnamefont {Reed}}, \bibinfo {author} {\bibfnamefont {L.}~\bibnamefont {Vivien}}, \bibinfo {author} {\bibfnamefont {D.}~\bibnamefont {{Marris-Morini}}}, \bibinfo {author} {\bibfnamefont {E.}~\bibnamefont {Cassan}}, \bibinfo {author} {\bibfnamefont {L.}~\bibnamefont {Virot}}, \bibinfo {author} {\bibfnamefont {J.-M.}\ \bibnamefont {F{\'e}d{\'e}li}}, \bibinfo {author} {\bibfnamefont {J.-M.}\ \bibnamefont {Hartmann}}, \bibinfo {author} {\bibfnamefont {J.~H.}\ \bibnamefont {Schmid}}, \bibinfo {author} {\bibfnamefont {D.-X.}\ \bibnamefont {Xu}}, \bibinfo {author} {\bibfnamefont {F.}~\bibnamefont {Boeuf}}, \bibinfo {author} {\bibfnamefont {P.}~\bibnamefont {O'Brien}}, \bibinfo {author} {\bibfnamefont {G.~Z.}\ \bibnamefont {Mashanovich}},\ and\ \bibinfo {author} {\bibfnamefont {M.}~\bibnamefont {Nedeljkovic}},\ }\href {https://doi.org/10.1088/2040-8978/18/7/073003} {\bibfield  {journal} {\bibinfo  {journal} {Journal of Optics}\ }\textbf {\bibinfo {volume} {18}},\ \bibinfo {pages} {073003} (\bibinfo {year} {2016})}\BibitemShut {NoStop}%
    \bibitem [{\citenamefont {{Sorace-Agaskar}}\ \emph {et~al.}(2019)\citenamefont {{Sorace-Agaskar}}, \citenamefont {Kharas}, \citenamefont {Yegnanarayanan}, \citenamefont {Maxson}, \citenamefont {West}, \citenamefont {Loh}, \citenamefont {Bramhavar}, \citenamefont {Ram}, \citenamefont {Chiaverini}, \citenamefont {Sage},\ and\ \citenamefont {Juodawlkis}}]{sorace-agaskar_versatile_2019}%
      \BibitemOpen
      \bibfield  {author} {\bibinfo {author} {\bibfnamefont {C.}~\bibnamefont {{Sorace-Agaskar}}}, \bibinfo {author} {\bibfnamefont {D.}~\bibnamefont {Kharas}}, \bibinfo {author} {\bibfnamefont {S.}~\bibnamefont {Yegnanarayanan}}, \bibinfo {author} {\bibfnamefont {R.~T.}\ \bibnamefont {Maxson}}, \bibinfo {author} {\bibfnamefont {G.~N.}\ \bibnamefont {West}}, \bibinfo {author} {\bibfnamefont {W.}~\bibnamefont {Loh}}, \bibinfo {author} {\bibfnamefont {S.}~\bibnamefont {Bramhavar}}, \bibinfo {author} {\bibfnamefont {R.~J.}\ \bibnamefont {Ram}}, \bibinfo {author} {\bibfnamefont {J.}~\bibnamefont {Chiaverini}}, \bibinfo {author} {\bibfnamefont {J.}~\bibnamefont {Sage}},\ and\ \bibinfo {author} {\bibfnamefont {P.}~\bibnamefont {Juodawlkis}},\ }\href {https://doi.org/10.1109/JSTQE.2019.2904443} {\bibfield  {journal} {\bibinfo  {journal} {IEEE Journal of Selected Topics in Quantum Electronics}\ }\textbf {\bibinfo {volume} {25}},\ \bibinfo {pages} {1} (\bibinfo {year} {2019})}\BibitemShut {NoStop}%
    \bibitem [{\citenamefont {Blumenthal}(2020)}]{blumenthal_photonic_2020}%
      \BibitemOpen
      \bibfield  {author} {\bibinfo {author} {\bibfnamefont {D.~J.}\ \bibnamefont {Blumenthal}},\ }\href {https://doi.org/10.1063/1.5131683} {\bibfield  {journal} {\bibinfo  {journal} {APL Photonics}\ }\textbf {\bibinfo {volume} {5}},\ \bibinfo {pages} {020903} (\bibinfo {year} {2020})}\BibitemShut {NoStop}%
    \bibitem [{\citenamefont {Ropp}\ \emph {et~al.}(2023)\citenamefont {Ropp}, \citenamefont {Zhu}, \citenamefont {Yulaev}, \citenamefont {Westly}, \citenamefont {Simelgor}, \citenamefont {Rakholia}, \citenamefont {Lunden}, \citenamefont {Sheredy}, \citenamefont {Boyd}, \citenamefont {Papp}, \citenamefont {Agrawal},\ and\ \citenamefont {Aksyuk}}]{RoppLightSciAppl2023}%
      \BibitemOpen
      \bibfield  {author} {\bibinfo {author} {\bibfnamefont {C.}~\bibnamefont {Ropp}}, \bibinfo {author} {\bibfnamefont {W.}~\bibnamefont {Zhu}}, \bibinfo {author} {\bibfnamefont {A.}~\bibnamefont {Yulaev}}, \bibinfo {author} {\bibfnamefont {D.}~\bibnamefont {Westly}}, \bibinfo {author} {\bibfnamefont {G.}~\bibnamefont {Simelgor}}, \bibinfo {author} {\bibfnamefont {A.}~\bibnamefont {Rakholia}}, \bibinfo {author} {\bibfnamefont {W.}~\bibnamefont {Lunden}}, \bibinfo {author} {\bibfnamefont {D.}~\bibnamefont {Sheredy}}, \bibinfo {author} {\bibfnamefont {M.~M.}\ \bibnamefont {Boyd}}, \bibinfo {author} {\bibfnamefont {S.}~\bibnamefont {Papp}}, \bibinfo {author} {\bibfnamefont {A.}~\bibnamefont {Agrawal}},\ and\ \bibinfo {author} {\bibfnamefont {V.}~\bibnamefont {Aksyuk}},\ }\href {https://doi.org/10.1038/s41377-023-01081-x} {\bibfield  {journal} {\bibinfo  {journal} {Light: Science \& Applications}\ }\textbf {\bibinfo {volume} {12}},\ \bibinfo {pages} {83} (\bibinfo {year} {2023})}\BibitemShut {NoStop}%
    \bibitem [{\citenamefont {Lu}\ \emph {et~al.}(2019{\natexlab{a}})\citenamefont {Lu}, \citenamefont {Li}, \citenamefont {Westly}, \citenamefont {Moille}, \citenamefont {Singh}, \citenamefont {Anant},\ and\ \citenamefont {Srinivasan}}]{LuNat.Phys.2019}%
      \BibitemOpen
      \bibfield  {author} {\bibinfo {author} {\bibfnamefont {X.}~\bibnamefont {Lu}}, \bibinfo {author} {\bibfnamefont {Q.}~\bibnamefont {Li}}, \bibinfo {author} {\bibfnamefont {D.~A.}\ \bibnamefont {Westly}}, \bibinfo {author} {\bibfnamefont {G.}~\bibnamefont {Moille}}, \bibinfo {author} {\bibfnamefont {A.}~\bibnamefont {Singh}}, \bibinfo {author} {\bibfnamefont {V.}~\bibnamefont {Anant}},\ and\ \bibinfo {author} {\bibfnamefont {K.}~\bibnamefont {Srinivasan}},\ }\href {https://doi.org/10.1038/s41567-018-0394-3} {\bibfield  {journal} {\bibinfo  {journal} {Nature physics}\ }\textbf {\bibinfo {volume} {15}},\ \bibinfo {pages} {373} (\bibinfo {year} {2019}{\natexlab{a}})}\BibitemShut {NoStop}%
    \bibitem [{\citenamefont {Sinclair}\ \emph {et~al.}(2020{\natexlab{a}})\citenamefont {Sinclair}, \citenamefont {Gallacher}, \citenamefont {Sorel}, \citenamefont {Bayley}, \citenamefont {McBrearty}, \citenamefont {Millar}, \citenamefont {Hild},\ and\ \citenamefont {Paul}}]{SinclairOpt.ExpressOE2020a}%
      \BibitemOpen
      \bibfield  {author} {\bibinfo {author} {\bibfnamefont {M.}~\bibnamefont {Sinclair}}, \bibinfo {author} {\bibfnamefont {K.}~\bibnamefont {Gallacher}}, \bibinfo {author} {\bibfnamefont {M.}~\bibnamefont {Sorel}}, \bibinfo {author} {\bibfnamefont {J.~C.}\ \bibnamefont {Bayley}}, \bibinfo {author} {\bibfnamefont {E.}~\bibnamefont {McBrearty}}, \bibinfo {author} {\bibfnamefont {R.~W.}\ \bibnamefont {Millar}}, \bibinfo {author} {\bibfnamefont {S.}~\bibnamefont {Hild}},\ and\ \bibinfo {author} {\bibfnamefont {D.~J.}\ \bibnamefont {Paul}},\ }\href {https://doi.org/10.1364/OE.381224} {\bibfield  {journal} {\bibinfo  {journal} {Optics Express}\ }\textbf {\bibinfo {volume} {28}},\ \bibinfo {pages} {4010} (\bibinfo {year} {2020}{\natexlab{a}})}\BibitemShut {NoStop}%
    \bibitem [{\citenamefont {Martin}\ \emph {et~al.}(2017)\citenamefont {Martin}, \citenamefont {Combri{\'e}},\ and\ \citenamefont {Rossi}}]{MartinJ.Opt.2017}%
      \BibitemOpen
      \bibfield  {author} {\bibinfo {author} {\bibfnamefont {A.}~\bibnamefont {Martin}}, \bibinfo {author} {\bibfnamefont {S.}~\bibnamefont {Combri{\'e}}},\ and\ \bibinfo {author} {\bibfnamefont {A.~D.}\ \bibnamefont {Rossi}},\ }\href {https://doi.org/10.1088/2040-8986/aa5498} {\bibfield  {journal} {\bibinfo  {journal} {Journal of Optics}\ }\textbf {\bibinfo {volume} {19}},\ \bibinfo {pages} {033002} (\bibinfo {year} {2017})}\BibitemShut {NoStop}%
    \bibitem [{\citenamefont {Marty}\ \emph {et~al.}(2021)\citenamefont {Marty}, \citenamefont {Combri{\'e}}, \citenamefont {Raineri},\ and\ \citenamefont {De~Rossi}}]{MartyNat.Photonics2021}%
      \BibitemOpen
      \bibfield  {author} {\bibinfo {author} {\bibfnamefont {G.}~\bibnamefont {Marty}}, \bibinfo {author} {\bibfnamefont {S.}~\bibnamefont {Combri{\'e}}}, \bibinfo {author} {\bibfnamefont {F.}~\bibnamefont {Raineri}},\ and\ \bibinfo {author} {\bibfnamefont {A.}~\bibnamefont {De~Rossi}},\ }\href {https://doi.org/10.1038/s41566-020-00737-z} {\bibfield  {journal} {\bibinfo  {journal} {Nature Photonics}\ }\textbf {\bibinfo {volume} {15}},\ \bibinfo {pages} {53} (\bibinfo {year} {2021})}\BibitemShut {NoStop}%
    \bibitem [{\citenamefont {Lu}\ \emph {et~al.}(2019{\natexlab{b}})\citenamefont {Lu}, \citenamefont {Moille}, \citenamefont {Singh}, \citenamefont {Li}, \citenamefont {Westly}, \citenamefont {Rao}, \citenamefont {Yu}, \citenamefont {Briles}, \citenamefont {Papp},\ and\ \citenamefont {Srinivasan}}]{LuOptica2019}%
      \BibitemOpen
      \bibfield  {author} {\bibinfo {author} {\bibfnamefont {X.}~\bibnamefont {Lu}}, \bibinfo {author} {\bibfnamefont {G.}~\bibnamefont {Moille}}, \bibinfo {author} {\bibfnamefont {A.}~\bibnamefont {Singh}}, \bibinfo {author} {\bibfnamefont {Q.}~\bibnamefont {Li}}, \bibinfo {author} {\bibfnamefont {D.~A.}\ \bibnamefont {Westly}}, \bibinfo {author} {\bibfnamefont {A.}~\bibnamefont {Rao}}, \bibinfo {author} {\bibfnamefont {S.-P.}\ \bibnamefont {Yu}}, \bibinfo {author} {\bibfnamefont {T.~C.}\ \bibnamefont {Briles}}, \bibinfo {author} {\bibfnamefont {S.~B.}\ \bibnamefont {Papp}},\ and\ \bibinfo {author} {\bibfnamefont {K.}~\bibnamefont {Srinivasan}},\ }\href {https://doi.org/10.1364/OPTICA.6.001535} {\bibfield  {journal} {\bibinfo  {journal} {Optica}\ }\textbf {\bibinfo {volume} {6}},\ \bibinfo {pages} {1535} (\bibinfo {year} {2019}{\natexlab{b}})}\BibitemShut {NoStop}%
    \bibitem [{\citenamefont {Stone}\ \emph {et~al.}(2023)\citenamefont {Stone}, \citenamefont {Lu}, \citenamefont {Moille}, \citenamefont {Westly}, \citenamefont {Rahman},\ and\ \citenamefont {Srinivasan}}]{StoneNat.Photon.2023}%
      \BibitemOpen
      \bibfield  {author} {\bibinfo {author} {\bibfnamefont {J.~R.}\ \bibnamefont {Stone}}, \bibinfo {author} {\bibfnamefont {X.}~\bibnamefont {Lu}}, \bibinfo {author} {\bibfnamefont {G.}~\bibnamefont {Moille}}, \bibinfo {author} {\bibfnamefont {D.}~\bibnamefont {Westly}}, \bibinfo {author} {\bibfnamefont {T.}~\bibnamefont {Rahman}},\ and\ \bibinfo {author} {\bibfnamefont {K.}~\bibnamefont {Srinivasan}},\ }\href {https://doi.org/10.1038/s41566-023-01326-6} {\bibfield  {journal} {\bibinfo  {journal} {Nature Photonics}\ ,\ \bibinfo {pages} {1}} (\bibinfo {year} {2023})}\BibitemShut {NoStop}%
    \bibitem [{\citenamefont {Newman}\ \emph {et~al.}(2019)\citenamefont {Newman}, \citenamefont {Maurice}, \citenamefont {Drake}, \citenamefont {Stone}, \citenamefont {Briles}, \citenamefont {Spencer}, \citenamefont {Fredrick}, \citenamefont {Li}, \citenamefont {Westly}, \citenamefont {Ilic}, \citenamefont {Shen}, \citenamefont {Suh}, \citenamefont {Yang}, \citenamefont {Johnson}, \citenamefont {Johnson}, \citenamefont {Hollberg}, \citenamefont {Vahala}, \citenamefont {Srinivasan}, \citenamefont {Diddams}, \citenamefont {Kitching}, \citenamefont {Papp},\ and\ \citenamefont {Hummon}}]{NewmanOptica2019}%
      \BibitemOpen
      \bibfield  {author} {\bibinfo {author} {\bibfnamefont {Z.~L.}\ \bibnamefont {Newman}}, \bibinfo {author} {\bibfnamefont {V.}~\bibnamefont {Maurice}}, \bibinfo {author} {\bibfnamefont {T.}~\bibnamefont {Drake}}, \bibinfo {author} {\bibfnamefont {J.~R.}\ \bibnamefont {Stone}}, \bibinfo {author} {\bibfnamefont {T.~C.}\ \bibnamefont {Briles}}, \bibinfo {author} {\bibfnamefont {D.~T.}\ \bibnamefont {Spencer}}, \bibinfo {author} {\bibfnamefont {C.}~\bibnamefont {Fredrick}}, \bibinfo {author} {\bibfnamefont {Q.}~\bibnamefont {Li}}, \bibinfo {author} {\bibfnamefont {D.}~\bibnamefont {Westly}}, \bibinfo {author} {\bibfnamefont {B.~R.}\ \bibnamefont {Ilic}}, \bibinfo {author} {\bibfnamefont {B.}~\bibnamefont {Shen}}, \bibinfo {author} {\bibfnamefont {M.-G.}\ \bibnamefont {Suh}}, \bibinfo {author} {\bibfnamefont {K.~Y.}\ \bibnamefont {Yang}}, \bibinfo {author} {\bibfnamefont {C.}~\bibnamefont {Johnson}}, \bibinfo {author} {\bibfnamefont {D.~M.~S.}\ \bibnamefont {Johnson}}, \bibinfo {author} {\bibfnamefont {L.}~\bibnamefont {Hollberg}}, \bibinfo {author} {\bibfnamefont {K.~J.}\ \bibnamefont {Vahala}}, \bibinfo {author} {\bibfnamefont {K.}~\bibnamefont {Srinivasan}}, \bibinfo {author} {\bibfnamefont {S.~A.}\ \bibnamefont {Diddams}}, \bibinfo {author} {\bibfnamefont {J.}~\bibnamefont {Kitching}}, \bibinfo {author} {\bibfnamefont {S.~B.}\ \bibnamefont {Papp}},\ and\ \bibinfo {author} {\bibfnamefont {M.~T.}\ \bibnamefont {Hummon}},\ }\href {https://doi.org/10.1364/OPTICA.6.000680} {\bibfield  {journal} {\bibinfo  {journal} {Optica}\ }\textbf {\bibinfo {volume} {6}},\ \bibinfo {pages} {680} (\bibinfo {year} {2019})}\BibitemShut {NoStop}%
    \bibitem [{\citenamefont {Yu}\ \emph {et~al.}(2019)\citenamefont {Yu}, \citenamefont {Briles}, \citenamefont {Moille}, \citenamefont {Lu}, \citenamefont {Diddams}, \citenamefont {Srinivasan},\ and\ \citenamefont {Papp}}]{YuPhys.Rev.Applied2019a}%
      \BibitemOpen
      \bibfield  {author} {\bibinfo {author} {\bibfnamefont {S.-P.}\ \bibnamefont {Yu}}, \bibinfo {author} {\bibfnamefont {T.~C.}\ \bibnamefont {Briles}}, \bibinfo {author} {\bibfnamefont {G.~T.}\ \bibnamefont {Moille}}, \bibinfo {author} {\bibfnamefont {X.}~\bibnamefont {Lu}}, \bibinfo {author} {\bibfnamefont {S.~A.}\ \bibnamefont {Diddams}}, \bibinfo {author} {\bibfnamefont {K.}~\bibnamefont {Srinivasan}},\ and\ \bibinfo {author} {\bibfnamefont {S.~B.}\ \bibnamefont {Papp}},\ }\href {https://doi.org/10.1103/PhysRevApplied.11.044017} {\bibfield  {journal} {\bibinfo  {journal} {Physical Review Applied}\ }\textbf {\bibinfo {volume} {11}},\ \bibinfo {pages} {044017} (\bibinfo {year} {2019})}\BibitemShut {NoStop}%
    \bibitem [{\citenamefont {Moille}\ \emph {et~al.}(2023)\citenamefont {Moille}, \citenamefont {Stone}, \citenamefont {Chojnacky}, \citenamefont {Shrestha}, \citenamefont {Javid}, \citenamefont {Menyuk},\ and\ \citenamefont {Srinivasan}}]{MoilleNature2023}%
      \BibitemOpen
      \bibfield  {author} {\bibinfo {author} {\bibfnamefont {G.}~\bibnamefont {Moille}}, \bibinfo {author} {\bibfnamefont {J.}~\bibnamefont {Stone}}, \bibinfo {author} {\bibfnamefont {M.}~\bibnamefont {Chojnacky}}, \bibinfo {author} {\bibfnamefont {R.}~\bibnamefont {Shrestha}}, \bibinfo {author} {\bibfnamefont {U.~A.}\ \bibnamefont {Javid}}, \bibinfo {author} {\bibfnamefont {C.}~\bibnamefont {Menyuk}},\ and\ \bibinfo {author} {\bibfnamefont {K.}~\bibnamefont {Srinivasan}},\ }\href {https://doi.org/10.1038/s41586-023-06730-0} {\bibfield  {journal} {\bibinfo  {journal} {Nature}\ }\textbf {\bibinfo {volume} {624}},\ \bibinfo {pages} {267} (\bibinfo {year} {2023})}\BibitemShut {NoStop}%
    \bibitem [{\citenamefont {Long}\ \emph {et~al.}(2023)\citenamefont {Long}, \citenamefont {Stone}, \citenamefont {Sun}, \citenamefont {Westly},\ and\ \citenamefont {Srinivasan}}]{Long2023}%
      \BibitemOpen
      \bibfield  {author} {\bibinfo {author} {\bibfnamefont {D.~A.}\ \bibnamefont {Long}}, \bibinfo {author} {\bibfnamefont {J.~R.}\ \bibnamefont {Stone}}, \bibinfo {author} {\bibfnamefont {Y.}~\bibnamefont {Sun}}, \bibinfo {author} {\bibfnamefont {D.}~\bibnamefont {Westly}},\ and\ \bibinfo {author} {\bibfnamefont {K.}~\bibnamefont {Srinivasan}},\ }\href {https://doi.org/10.48550/arXiv.2309.16069} {\bibinfo {title} {Sub-{{Doppler}} spectroscopy of quantum systems through nanophotonic spectral translation of electro-optic light}} (\bibinfo {year} {2023}),\ \Eprint {https://arxiv.org/abs/2309.16069} {arxiv:2309.16069 [physics]} \BibitemShut {NoStop}%
    \bibitem [{\citenamefont {Sacher}\ \emph {et~al.}(2019)\citenamefont {Sacher}, \citenamefont {Luo}, \citenamefont {Yang}, \citenamefont {Chen}, \citenamefont {Lordello}, \citenamefont {Mak}, \citenamefont {Liu}, \citenamefont {Hu}, \citenamefont {Xue}, \citenamefont {{Guo-Qiang Lo}}, \citenamefont {Roukes},\ and\ \citenamefont {Poon}}]{sacher_visible-light_2019}%
      \BibitemOpen
      \bibfield  {author} {\bibinfo {author} {\bibfnamefont {W.~D.}\ \bibnamefont {Sacher}}, \bibinfo {author} {\bibfnamefont {X.}~\bibnamefont {Luo}}, \bibinfo {author} {\bibfnamefont {Y.}~\bibnamefont {Yang}}, \bibinfo {author} {\bibfnamefont {F.-D.}\ \bibnamefont {Chen}}, \bibinfo {author} {\bibfnamefont {T.}~\bibnamefont {Lordello}}, \bibinfo {author} {\bibfnamefont {J.~C.~C.}\ \bibnamefont {Mak}}, \bibinfo {author} {\bibfnamefont {X.}~\bibnamefont {Liu}}, \bibinfo {author} {\bibfnamefont {T.}~\bibnamefont {Hu}}, \bibinfo {author} {\bibfnamefont {T.}~\bibnamefont {Xue}}, \bibinfo {author} {\bibfnamefont {P.}~\bibnamefont {{Guo-Qiang Lo}}}, \bibinfo {author} {\bibfnamefont {M.~L.}\ \bibnamefont {Roukes}},\ and\ \bibinfo {author} {\bibfnamefont {J.~K.~S.}\ \bibnamefont {Poon}},\ }\href {https://doi.org/10.1364/OE.27.037400} {\bibfield  {journal} {\bibinfo  {journal} {Optics Express}\ }\textbf {\bibinfo {volume} {27}},\ \bibinfo {pages} {37400} (\bibinfo {year} {2019})}\BibitemShut {NoStop}%
    \bibitem [{\citenamefont {Pfeiffer}\ \emph {et~al.}(2018)\citenamefont {Pfeiffer}, \citenamefont {Liu}, \citenamefont {Raja}, \citenamefont {Morais}, \citenamefont {Ghadiani},\ and\ \citenamefont {Kippenberg}}]{PfeifferOpticaOPTICA2018}%
      \BibitemOpen
      \bibfield  {author} {\bibinfo {author} {\bibfnamefont {M.~H.~P.}\ \bibnamefont {Pfeiffer}}, \bibinfo {author} {\bibfnamefont {J.}~\bibnamefont {Liu}}, \bibinfo {author} {\bibfnamefont {A.~S.}\ \bibnamefont {Raja}}, \bibinfo {author} {\bibfnamefont {T.}~\bibnamefont {Morais}}, \bibinfo {author} {\bibfnamefont {B.}~\bibnamefont {Ghadiani}},\ and\ \bibinfo {author} {\bibfnamefont {T.~J.}\ \bibnamefont {Kippenberg}},\ }\href {https://doi.org/10.1364/OPTICA.5.000884} {\bibfield  {journal} {\bibinfo  {journal} {Optica}\ }\textbf {\bibinfo {volume} {5}},\ \bibinfo {pages} {884} (\bibinfo {year} {2018})}\BibitemShut {NoStop}%
    \bibitem [{\citenamefont {Ji}\ \emph {et~al.}(2021)\citenamefont {Ji}, \citenamefont {Roberts}, \citenamefont {{Corato-Zanarella}},\ and\ \citenamefont {Lipson}}]{ji_methods_2021}%
      \BibitemOpen
      \bibfield  {author} {\bibinfo {author} {\bibfnamefont {X.}~\bibnamefont {Ji}}, \bibinfo {author} {\bibfnamefont {S.}~\bibnamefont {Roberts}}, \bibinfo {author} {\bibfnamefont {M.}~\bibnamefont {{Corato-Zanarella}}},\ and\ \bibinfo {author} {\bibfnamefont {M.}~\bibnamefont {Lipson}},\ }\href {https://doi.org/10.1063/5.0057881} {\bibfield  {journal} {\bibinfo  {journal} {APL Photonics}\ }\textbf {\bibinfo {volume} {6}},\ \bibinfo {pages} {071101} (\bibinfo {year} {2021})}\BibitemShut {NoStop}%
    \bibitem [{\citenamefont {Liu}\ \emph {et~al.}(2021{\natexlab{a}})\citenamefont {Liu}, \citenamefont {Huang}, \citenamefont {Wang}, \citenamefont {He}, \citenamefont {Raja}, \citenamefont {Liu}, \citenamefont {Engelsen},\ and\ \citenamefont {Kippenberg}}]{LiuNatCommun2021a}%
      \BibitemOpen
      \bibfield  {author} {\bibinfo {author} {\bibfnamefont {J.}~\bibnamefont {Liu}}, \bibinfo {author} {\bibfnamefont {G.}~\bibnamefont {Huang}}, \bibinfo {author} {\bibfnamefont {R.~N.}\ \bibnamefont {Wang}}, \bibinfo {author} {\bibfnamefont {J.}~\bibnamefont {He}}, \bibinfo {author} {\bibfnamefont {A.~S.}\ \bibnamefont {Raja}}, \bibinfo {author} {\bibfnamefont {T.}~\bibnamefont {Liu}}, \bibinfo {author} {\bibfnamefont {N.~J.}\ \bibnamefont {Engelsen}},\ and\ \bibinfo {author} {\bibfnamefont {T.~J.}\ \bibnamefont {Kippenberg}},\ }\href {https://doi.org/10.1038/s41467-021-21973-z} {\bibfield  {journal} {\bibinfo  {journal} {Nature Communications}\ }\textbf {\bibinfo {volume} {12}},\ \bibinfo {pages} {2236} (\bibinfo {year} {2021}{\natexlab{a}})}\BibitemShut {NoStop}%
    \bibitem [{\citenamefont {Chang}\ \emph {et~al.}(2020)\citenamefont {Chang}, \citenamefont {Xie}, \citenamefont {Shu}, \citenamefont {Yang}, \citenamefont {Shen}, \citenamefont {Boes}, \citenamefont {Peters}, \citenamefont {Jin}, \citenamefont {Xiang}, \citenamefont {Liu} \emph {et~al.}}]{ChangNat.Commun.2020}%
      \BibitemOpen
      \bibfield  {author} {\bibinfo {author} {\bibfnamefont {L.}~\bibnamefont {Chang}}, \bibinfo {author} {\bibfnamefont {W.}~\bibnamefont {Xie}}, \bibinfo {author} {\bibfnamefont {H.}~\bibnamefont {Shu}}, \bibinfo {author} {\bibfnamefont {Q.-F.}\ \bibnamefont {Yang}}, \bibinfo {author} {\bibfnamefont {B.}~\bibnamefont {Shen}}, \bibinfo {author} {\bibfnamefont {A.}~\bibnamefont {Boes}}, \bibinfo {author} {\bibfnamefont {J.~D.}\ \bibnamefont {Peters}}, \bibinfo {author} {\bibfnamefont {W.}~\bibnamefont {Jin}}, \bibinfo {author} {\bibfnamefont {C.}~\bibnamefont {Xiang}}, \bibinfo {author} {\bibfnamefont {S.}~\bibnamefont {Liu}}, \emph {et~al.},\ }\href {https://doi.org/10.1038/s41467-020-15005-5} {\bibfield  {journal} {\bibinfo  {journal} {Nature communications}\ }\textbf {\bibinfo {volume} {11}},\ \bibinfo {pages} {1} (\bibinfo {year} {2020})}\BibitemShut {NoStop}%
    \bibitem [{\citenamefont {Moille}\ \emph {et~al.}(2021{\natexlab{a}})\citenamefont {Moille}, \citenamefont {Westly}, \citenamefont {Orji},\ and\ \citenamefont {Srinivasan}}]{MoilleAppl.Phys.Lett.2021a}%
      \BibitemOpen
      \bibfield  {author} {\bibinfo {author} {\bibfnamefont {G.}~\bibnamefont {Moille}}, \bibinfo {author} {\bibfnamefont {D.}~\bibnamefont {Westly}}, \bibinfo {author} {\bibfnamefont {N.~G.}\ \bibnamefont {Orji}},\ and\ \bibinfo {author} {\bibfnamefont {K.}~\bibnamefont {Srinivasan}},\ }\href {https://doi.org/10.1063/5.0061238} {\bibfield  {journal} {\bibinfo  {journal} {Applied Physics Letters}\ }\textbf {\bibinfo {volume} {119}},\ \bibinfo {pages} {121103} (\bibinfo {year} {2021}{\natexlab{a}})}\BibitemShut {NoStop}%
    \bibitem [{\citenamefont {Moille}\ \emph {et~al.}(2021{\natexlab{b}})\citenamefont {Moille}, \citenamefont {Perez}, \citenamefont {Stone}, \citenamefont {Rao}, \citenamefont {Lu}, \citenamefont {Rahman}, \citenamefont {Chembo},\ and\ \citenamefont {Srinivasan}}]{MoilleNat.Commun.2021a}%
      \BibitemOpen
      \bibfield  {author} {\bibinfo {author} {\bibfnamefont {G.}~\bibnamefont {Moille}}, \bibinfo {author} {\bibfnamefont {E.~F.}\ \bibnamefont {Perez}}, \bibinfo {author} {\bibfnamefont {J.~R.}\ \bibnamefont {Stone}}, \bibinfo {author} {\bibfnamefont {A.}~\bibnamefont {Rao}}, \bibinfo {author} {\bibfnamefont {X.}~\bibnamefont {Lu}}, \bibinfo {author} {\bibfnamefont {T.~S.}\ \bibnamefont {Rahman}}, \bibinfo {author} {\bibfnamefont {Y.~K.}\ \bibnamefont {Chembo}},\ and\ \bibinfo {author} {\bibfnamefont {K.}~\bibnamefont {Srinivasan}},\ }\href {https://doi.org/10.1038/s41467-021-27469-0} {\bibfield  {journal} {\bibinfo  {journal} {Nature Communications}\ }\textbf {\bibinfo {volume} {12}},\ \bibinfo {pages} {7275} (\bibinfo {year} {2021}{\natexlab{b}})}\BibitemShut {NoStop}%
    \bibitem [{\citenamefont {Liu}\ \emph {et~al.}(2021{\natexlab{b}})\citenamefont {Liu}, \citenamefont {Gong}, \citenamefont {Bruch}, \citenamefont {Surya}, \citenamefont {Lu},\ and\ \citenamefont {Tang}}]{liu2021aluminum}%
      \BibitemOpen
      \bibfield  {author} {\bibinfo {author} {\bibfnamefont {X.}~\bibnamefont {Liu}}, \bibinfo {author} {\bibfnamefont {Z.}~\bibnamefont {Gong}}, \bibinfo {author} {\bibfnamefont {A.~W.}\ \bibnamefont {Bruch}}, \bibinfo {author} {\bibfnamefont {J.~B.}\ \bibnamefont {Surya}}, \bibinfo {author} {\bibfnamefont {J.}~\bibnamefont {Lu}},\ and\ \bibinfo {author} {\bibfnamefont {H.~X.}\ \bibnamefont {Tang}},\ }\href {https://doi.org/10.1038/s41467-021-25751-9} {\bibfield  {journal} {\bibinfo  {journal} {Nature communications}\ }\textbf {\bibinfo {volume} {12}},\ \bibinfo {pages} {5428} (\bibinfo {year} {2021}{\natexlab{b}})}\BibitemShut {NoStop}%
    \bibitem [{\citenamefont {Hansson}\ \emph {et~al.}(2013)\citenamefont {Hansson}, \citenamefont {Modotto},\ and\ \citenamefont {Wabnitz}}]{HanssonPhys.Rev.A2013b}%
      \BibitemOpen
      \bibfield  {author} {\bibinfo {author} {\bibfnamefont {T.}~\bibnamefont {Hansson}}, \bibinfo {author} {\bibfnamefont {D.}~\bibnamefont {Modotto}},\ and\ \bibinfo {author} {\bibfnamefont {S.}~\bibnamefont {Wabnitz}},\ }\href {https://doi.org/10.1103/PhysRevA.88.023819} {\bibfield  {journal} {\bibinfo  {journal} {Physical Review A}\ }\textbf {\bibinfo {volume} {88}},\ \bibinfo {pages} {023819} (\bibinfo {year} {2013})}\BibitemShut {NoStop}%
    \bibitem [{\citenamefont {Liu}\ \emph {et~al.}(2014)\citenamefont {Liu}, \citenamefont {Xuan}, \citenamefont {Xue}, \citenamefont {Wang}, \citenamefont {Chen}, \citenamefont {Metcalf}, \citenamefont {Wang}, \citenamefont {Leaird}, \citenamefont {Qi},\ and\ \citenamefont {Weiner}}]{liu2014investigation}%
      \BibitemOpen
      \bibfield  {author} {\bibinfo {author} {\bibfnamefont {Y.}~\bibnamefont {Liu}}, \bibinfo {author} {\bibfnamefont {Y.}~\bibnamefont {Xuan}}, \bibinfo {author} {\bibfnamefont {X.}~\bibnamefont {Xue}}, \bibinfo {author} {\bibfnamefont {P.-H.}\ \bibnamefont {Wang}}, \bibinfo {author} {\bibfnamefont {S.}~\bibnamefont {Chen}}, \bibinfo {author} {\bibfnamefont {A.~J.}\ \bibnamefont {Metcalf}}, \bibinfo {author} {\bibfnamefont {J.}~\bibnamefont {Wang}}, \bibinfo {author} {\bibfnamefont {D.~E.}\ \bibnamefont {Leaird}}, \bibinfo {author} {\bibfnamefont {M.}~\bibnamefont {Qi}},\ and\ \bibinfo {author} {\bibfnamefont {A.~M.}\ \bibnamefont {Weiner}},\ }\href {https://doi.org/10.1364/OPTICA.1.000137} {\bibfield  {journal} {\bibinfo  {journal} {optica}\ }\textbf {\bibinfo {volume} {1}},\ \bibinfo {pages} {137} (\bibinfo {year} {2014})}\BibitemShut {NoStop}%
    \bibitem [{\citenamefont {Xu}\ \emph {et~al.}(2021)\citenamefont {Xu}, \citenamefont {Sharples}, \citenamefont {Fatome}, \citenamefont {Coen}, \citenamefont {Erkintalo},\ and\ \citenamefont {Murdoch}}]{xu2021frequency}%
      \BibitemOpen
      \bibfield  {author} {\bibinfo {author} {\bibfnamefont {Y.}~\bibnamefont {Xu}}, \bibinfo {author} {\bibfnamefont {A.}~\bibnamefont {Sharples}}, \bibinfo {author} {\bibfnamefont {J.}~\bibnamefont {Fatome}}, \bibinfo {author} {\bibfnamefont {S.}~\bibnamefont {Coen}}, \bibinfo {author} {\bibfnamefont {M.}~\bibnamefont {Erkintalo}},\ and\ \bibinfo {author} {\bibfnamefont {S.~G.}\ \bibnamefont {Murdoch}},\ }\href {https://doi.org/10.1364/OL.413585} {\bibfield  {journal} {\bibinfo  {journal} {Optics Letters}\ }\textbf {\bibinfo {volume} {46}},\ \bibinfo {pages} {512} (\bibinfo {year} {2021})}\BibitemShut {NoStop}%
    \bibitem [{\citenamefont {Yu}\ \emph {et~al.}(2021)\citenamefont {Yu}, \citenamefont {Cole}, \citenamefont {Jung}, \citenamefont {Moille}, \citenamefont {Srinivasan},\ and\ \citenamefont {Papp}}]{YuNat.Photonics2021}%
      \BibitemOpen
      \bibfield  {author} {\bibinfo {author} {\bibfnamefont {S.-P.}\ \bibnamefont {Yu}}, \bibinfo {author} {\bibfnamefont {D.~C.}\ \bibnamefont {Cole}}, \bibinfo {author} {\bibfnamefont {H.}~\bibnamefont {Jung}}, \bibinfo {author} {\bibfnamefont {G.~T.}\ \bibnamefont {Moille}}, \bibinfo {author} {\bibfnamefont {K.}~\bibnamefont {Srinivasan}},\ and\ \bibinfo {author} {\bibfnamefont {S.~B.}\ \bibnamefont {Papp}},\ }\href {https://doi.org/10.1038/s41566-021-00800-3} {\bibfield  {journal} {\bibinfo  {journal} {Nature Photonics}\ }\textbf {\bibinfo {volume} {15}},\ \bibinfo {pages} {461} (\bibinfo {year} {2021})}\BibitemShut {NoStop}%
    \bibitem [{\citenamefont {Moille}\ \emph {et~al.}(2022{\natexlab{a}})\citenamefont {Moille}, \citenamefont {Westly}, \citenamefont {Simelgor},\ and\ \citenamefont {Srinivasan}}]{moille2022towards}%
      \BibitemOpen
      \bibfield  {author} {\bibinfo {author} {\bibfnamefont {G.}~\bibnamefont {Moille}}, \bibinfo {author} {\bibfnamefont {D.}~\bibnamefont {Westly}}, \bibinfo {author} {\bibfnamefont {G.}~\bibnamefont {Simelgor}},\ and\ \bibinfo {author} {\bibfnamefont {K.}~\bibnamefont {Srinivasan}},\ }in\ \href@noop {} {\emph {\bibinfo {booktitle} {{{CLEO}}: {{Science}} and Innovations}}}\ (\bibinfo {organization} {Optica Publishing Group},\ \bibinfo {year} {2022})\ pp.\ \bibinfo {pages} {SW4H--6}\BibitemShut {NoStop}%
    \bibitem [{\citenamefont {Li}\ \emph {et~al.}(2017)\citenamefont {Li}, \citenamefont {Briles}, \citenamefont {Westly}, \citenamefont {Drake}, \citenamefont {Stone}, \citenamefont {Ilic}, \citenamefont {Diddams}, \citenamefont {Papp},\ and\ \citenamefont {Srinivasan}}]{LiOptica2017a}%
      \BibitemOpen
      \bibfield  {author} {\bibinfo {author} {\bibfnamefont {Q.}~\bibnamefont {Li}}, \bibinfo {author} {\bibfnamefont {T.~C.}\ \bibnamefont {Briles}}, \bibinfo {author} {\bibfnamefont {D.~A.}\ \bibnamefont {Westly}}, \bibinfo {author} {\bibfnamefont {T.~E.}\ \bibnamefont {Drake}}, \bibinfo {author} {\bibfnamefont {J.~R.}\ \bibnamefont {Stone}}, \bibinfo {author} {\bibfnamefont {B.~R.}\ \bibnamefont {Ilic}}, \bibinfo {author} {\bibfnamefont {S.~A.}\ \bibnamefont {Diddams}}, \bibinfo {author} {\bibfnamefont {S.~B.}\ \bibnamefont {Papp}},\ and\ \bibinfo {author} {\bibfnamefont {K.}~\bibnamefont {Srinivasan}},\ }\href {https://doi.org/10.1364/OPTICA.4.000193} {\bibfield  {journal} {\bibinfo  {journal} {Optica}\ }\textbf {\bibinfo {volume} {4}},\ \bibinfo {pages} {193} (\bibinfo {year} {2017})}\BibitemShut {NoStop}%
    \bibitem [{\citenamefont {Brasch}\ \emph {et~al.}(2016)\citenamefont {Brasch}, \citenamefont {Geiselmann}, \citenamefont {Herr}, \citenamefont {Lihachev}, \citenamefont {Pfeiffer}, \citenamefont {Gorodetsky},\ and\ \citenamefont {Kippenberg}}]{BraschScience2016}%
      \BibitemOpen
      \bibfield  {author} {\bibinfo {author} {\bibfnamefont {V.}~\bibnamefont {Brasch}}, \bibinfo {author} {\bibfnamefont {M.}~\bibnamefont {Geiselmann}}, \bibinfo {author} {\bibfnamefont {T.}~\bibnamefont {Herr}}, \bibinfo {author} {\bibfnamefont {G.}~\bibnamefont {Lihachev}}, \bibinfo {author} {\bibfnamefont {M.~H.}\ \bibnamefont {Pfeiffer}}, \bibinfo {author} {\bibfnamefont {M.~L.}\ \bibnamefont {Gorodetsky}},\ and\ \bibinfo {author} {\bibfnamefont {T.~J.}\ \bibnamefont {Kippenberg}},\ }\href {https://doi.org/10.1126/science.aad4811} {\bibfield  {journal} {\bibinfo  {journal} {Science}\ }\textbf {\bibinfo {volume} {351}},\ \bibinfo {pages} {357} (\bibinfo {year} {2016})}\BibitemShut {NoStop}%
    \bibitem [{\citenamefont {Pfeiffer}\ \emph {et~al.}(2017)\citenamefont {Pfeiffer}, \citenamefont {Herkommer}, \citenamefont {Liu}, \citenamefont {Guo}, \citenamefont {Karpov}, \citenamefont {Lucas}, \citenamefont {Zervas},\ and\ \citenamefont {Kippenberg}}]{PfeifferOpticaOPTICA2017}%
      \BibitemOpen
      \bibfield  {author} {\bibinfo {author} {\bibfnamefont {M.~H.~P.}\ \bibnamefont {Pfeiffer}}, \bibinfo {author} {\bibfnamefont {C.}~\bibnamefont {Herkommer}}, \bibinfo {author} {\bibfnamefont {J.}~\bibnamefont {Liu}}, \bibinfo {author} {\bibfnamefont {H.}~\bibnamefont {Guo}}, \bibinfo {author} {\bibfnamefont {M.}~\bibnamefont {Karpov}}, \bibinfo {author} {\bibfnamefont {E.}~\bibnamefont {Lucas}}, \bibinfo {author} {\bibfnamefont {M.}~\bibnamefont {Zervas}},\ and\ \bibinfo {author} {\bibfnamefont {T.~J.}\ \bibnamefont {Kippenberg}},\ }\href {https://doi.org/10.1364/OPTICA.4.000684} {\bibfield  {journal} {\bibinfo  {journal} {Optica}\ }\textbf {\bibinfo {volume} {4}},\ \bibinfo {pages} {684} (\bibinfo {year} {2017})}\BibitemShut {NoStop}%
    \bibitem [{\citenamefont {Chembo}\ and\ \citenamefont {Menyuk}(2013)}]{ChemboPhys.Rev.A2013}%
      \BibitemOpen
      \bibfield  {author} {\bibinfo {author} {\bibfnamefont {Y.~K.}\ \bibnamefont {Chembo}}\ and\ \bibinfo {author} {\bibfnamefont {C.~R.}\ \bibnamefont {Menyuk}},\ }\href {https://doi.org/10.1103/PhysRevA.87.053852} {\bibfield  {journal} {\bibinfo  {journal} {Physical Review A}\ }\textbf {\bibinfo {volume} {87}},\ \bibinfo {pages} {053852} (\bibinfo {year} {2013})}\BibitemShut {NoStop}%
    \bibitem [{\citenamefont {Moille}\ \emph {et~al.}(2019{\natexlab{a}})\citenamefont {Moille}, \citenamefont {Li}, \citenamefont {Xiyuan},\ and\ \citenamefont {Srinivasan}}]{MoilleJ.RES.NATL.INST.STAN.2019}%
      \BibitemOpen
      \bibfield  {author} {\bibinfo {author} {\bibfnamefont {G.}~\bibnamefont {Moille}}, \bibinfo {author} {\bibfnamefont {Q.}~\bibnamefont {Li}}, \bibinfo {author} {\bibfnamefont {L.}~\bibnamefont {Xiyuan}},\ and\ \bibinfo {author} {\bibfnamefont {K.}~\bibnamefont {Srinivasan}},\ }\href {https://doi.org/10.6028/jres.124.012} {\bibfield  {journal} {\bibinfo  {journal} {Journal of Research of the NIST}\ }\textbf {\bibinfo {volume} {124}},\ \bibinfo {pages} {124012} (\bibinfo {year} {2019}{\natexlab{a}})}\BibitemShut {NoStop}%
    \bibitem [{\citenamefont {Moille}\ \emph {et~al.}(2021{\natexlab{c}})\citenamefont {Moille}, \citenamefont {Westly}, \citenamefont {Simelgor},\ and\ \citenamefont {Srinivasan}}]{MoilleOpt.Lett.OL2021a}%
      \BibitemOpen
      \bibfield  {author} {\bibinfo {author} {\bibfnamefont {G.}~\bibnamefont {Moille}}, \bibinfo {author} {\bibfnamefont {D.}~\bibnamefont {Westly}}, \bibinfo {author} {\bibfnamefont {G.}~\bibnamefont {Simelgor}},\ and\ \bibinfo {author} {\bibfnamefont {K.}~\bibnamefont {Srinivasan}},\ }\href {https://doi.org/10.1364/OL.440907} {\bibfield  {journal} {\bibinfo  {journal} {Optics Letters}\ }\textbf {\bibinfo {volume} {46}},\ \bibinfo {pages} {5970} (\bibinfo {year} {2021}{\natexlab{c}})}\BibitemShut {NoStop}%
    \bibitem [{\citenamefont {Sinclair}\ \emph {et~al.}(2020{\natexlab{b}})\citenamefont {Sinclair}, \citenamefont {Gallacher}, \citenamefont {Sorel}, \citenamefont {Bayley}, \citenamefont {McBrearty}, \citenamefont {Millar}, \citenamefont {Hild},\ and\ \citenamefont {Paul}}]{SinclairOpt.ExpressOE2020}%
      \BibitemOpen
      \bibfield  {author} {\bibinfo {author} {\bibfnamefont {M.}~\bibnamefont {Sinclair}}, \bibinfo {author} {\bibfnamefont {K.}~\bibnamefont {Gallacher}}, \bibinfo {author} {\bibfnamefont {M.}~\bibnamefont {Sorel}}, \bibinfo {author} {\bibfnamefont {J.~C.}\ \bibnamefont {Bayley}}, \bibinfo {author} {\bibfnamefont {E.}~\bibnamefont {McBrearty}}, \bibinfo {author} {\bibfnamefont {R.~W.}\ \bibnamefont {Millar}}, \bibinfo {author} {\bibfnamefont {S.}~\bibnamefont {Hild}},\ and\ \bibinfo {author} {\bibfnamefont {D.~J.}\ \bibnamefont {Paul}},\ }\href {https://doi.org/10.1364/OE.381224} {\bibfield  {journal} {\bibinfo  {journal} {Optics Express}\ }\textbf {\bibinfo {volume} {28}},\ \bibinfo {pages} {4010} (\bibinfo {year} {2020}{\natexlab{b}})}\BibitemShut {NoStop}%
    \bibitem [{\citenamefont {Lu}\ \emph {et~al.}(2020)\citenamefont {Lu}, \citenamefont {Moille}, \citenamefont {Rao}, \citenamefont {Westly},\ and\ \citenamefont {Srinivasan}}]{LuOpticaOPTICA2020}%
      \BibitemOpen
      \bibfield  {author} {\bibinfo {author} {\bibfnamefont {X.}~\bibnamefont {Lu}}, \bibinfo {author} {\bibfnamefont {G.}~\bibnamefont {Moille}}, \bibinfo {author} {\bibfnamefont {A.}~\bibnamefont {Rao}}, \bibinfo {author} {\bibfnamefont {D.~A.}\ \bibnamefont {Westly}},\ and\ \bibinfo {author} {\bibfnamefont {K.}~\bibnamefont {Srinivasan}},\ }\href {https://doi.org/10.1364/OPTICA.393810} {\bibfield  {journal} {\bibinfo  {journal} {Optica}\ }\textbf {\bibinfo {volume} {7}},\ \bibinfo {pages} {1417} (\bibinfo {year} {2020})}\BibitemShut {NoStop}%
    \bibitem [{\citenamefont {Moille}\ \emph {et~al.}(2019{\natexlab{b}})\citenamefont {Moille}, \citenamefont {Li}, \citenamefont {Briles}, \citenamefont {Yu}, \citenamefont {Drake}, \citenamefont {Lu}, \citenamefont {Rao}, \citenamefont {Westly}, \citenamefont {Papp},\ and\ \citenamefont {Srinivasan}}]{MoilleOpt.Lett.2019b}%
      \BibitemOpen
      \bibfield  {author} {\bibinfo {author} {\bibfnamefont {G.}~\bibnamefont {Moille}}, \bibinfo {author} {\bibfnamefont {Q.}~\bibnamefont {Li}}, \bibinfo {author} {\bibfnamefont {T.~C.}\ \bibnamefont {Briles}}, \bibinfo {author} {\bibfnamefont {S.-P.}\ \bibnamefont {Yu}}, \bibinfo {author} {\bibfnamefont {T.}~\bibnamefont {Drake}}, \bibinfo {author} {\bibfnamefont {X.}~\bibnamefont {Lu}}, \bibinfo {author} {\bibfnamefont {A.}~\bibnamefont {Rao}}, \bibinfo {author} {\bibfnamefont {D.}~\bibnamefont {Westly}}, \bibinfo {author} {\bibfnamefont {S.~B.}\ \bibnamefont {Papp}},\ and\ \bibinfo {author} {\bibfnamefont {K.}~\bibnamefont {Srinivasan}},\ }\href {https://doi.org/10.1364/OL.44.004737} {\bibfield  {journal} {\bibinfo  {journal} {Optics Letters}\ }\textbf {\bibinfo {volume} {44}},\ \bibinfo {pages} {4737} (\bibinfo {year} {2019}{\natexlab{b}})}\BibitemShut {NoStop}%
    \bibitem [{\citenamefont {Spencer}\ \emph {et~al.}(2018)\citenamefont {Spencer}, \citenamefont {Drake}, \citenamefont {Briles}, \citenamefont {Stone}, \citenamefont {Sinclair}, \citenamefont {Fredrick}, \citenamefont {Li}, \citenamefont {Westly}, \citenamefont {Ilic}, \citenamefont {Bluestone}, \citenamefont {Volet}, \citenamefont {Komljenovic}, \citenamefont {Chang}, \citenamefont {Lee}, \citenamefont {Oh}, \citenamefont {Suh}, \citenamefont {Yang}, \citenamefont {Pfeiffer}, \citenamefont {Kippenberg}, \citenamefont {Norberg}, \citenamefont {Theogarajan}, \citenamefont {Vahala}, \citenamefont {Newbury}, \citenamefont {Srinivasan}, \citenamefont {Bowers}, \citenamefont {Diddams},\ and\ \citenamefont {Papp}}]{SpencerNature2018}%
      \BibitemOpen
      \bibfield  {author} {\bibinfo {author} {\bibfnamefont {D.~T.}\ \bibnamefont {Spencer}}, \bibinfo {author} {\bibfnamefont {T.}~\bibnamefont {Drake}}, \bibinfo {author} {\bibfnamefont {T.~C.}\ \bibnamefont {Briles}}, \bibinfo {author} {\bibfnamefont {J.}~\bibnamefont {Stone}}, \bibinfo {author} {\bibfnamefont {L.~C.}\ \bibnamefont {Sinclair}}, \bibinfo {author} {\bibfnamefont {C.}~\bibnamefont {Fredrick}}, \bibinfo {author} {\bibfnamefont {Q.}~\bibnamefont {Li}}, \bibinfo {author} {\bibfnamefont {D.}~\bibnamefont {Westly}}, \bibinfo {author} {\bibfnamefont {B.~R.}\ \bibnamefont {Ilic}}, \bibinfo {author} {\bibfnamefont {A.}~\bibnamefont {Bluestone}}, \bibinfo {author} {\bibfnamefont {N.}~\bibnamefont {Volet}}, \bibinfo {author} {\bibfnamefont {T.}~\bibnamefont {Komljenovic}}, \bibinfo {author} {\bibfnamefont {L.}~\bibnamefont {Chang}}, \bibinfo {author} {\bibfnamefont {S.~H.}\ \bibnamefont {Lee}}, \bibinfo {author} {\bibfnamefont {D.~Y.}\ \bibnamefont {Oh}}, \bibinfo {author} {\bibfnamefont {M.-G.}\ \bibnamefont {Suh}}, \bibinfo {author} {\bibfnamefont {K.~Y.}\ \bibnamefont {Yang}}, \bibinfo {author} {\bibfnamefont {M.~H.~P.}\ \bibnamefont {Pfeiffer}}, \bibinfo {author} {\bibfnamefont {T.~J.}\ \bibnamefont {Kippenberg}}, \bibinfo {author} {\bibfnamefont {E.}~\bibnamefont {Norberg}}, \bibinfo {author} {\bibfnamefont {L.}~\bibnamefont {Theogarajan}}, \bibinfo {author} {\bibfnamefont {K.}~\bibnamefont {Vahala}}, \bibinfo {author} {\bibfnamefont {N.~R.}\ \bibnamefont {Newbury}}, \bibinfo {author} {\bibfnamefont {K.}~\bibnamefont {Srinivasan}}, \bibinfo {author} {\bibfnamefont {J.~E.}\ \bibnamefont {Bowers}}, \bibinfo {author} {\bibfnamefont {S.~A.}\ \bibnamefont {Diddams}},\ and\ \bibinfo {author} {\bibfnamefont {S.~B.}\ \bibnamefont {Papp}},\ }\href {https://doi.org/10.1038/s41586-018-0065-7} {\bibfield  {journal} {\bibinfo  {journal} {Nature}\ }\textbf {\bibinfo {volume} {557}},\ \bibinfo {pages} {81} (\bibinfo {year} {2018})}\BibitemShut {NoStop}%
    \bibitem [{\citenamefont {Zhao}\ \emph {et~al.}(2020)\citenamefont {Zhao}, \citenamefont {Ji}, \citenamefont {Kim}, \citenamefont {Donvalkar}, \citenamefont {Jang}, \citenamefont {Joshi}, \citenamefont {Yu}, \citenamefont {Joshi}, \citenamefont {Domeneguetti}, \citenamefont {Barbosa}, \citenamefont {Nussenzveig}, \citenamefont {Okawachi}, \citenamefont {Lipson},\ and\ \citenamefont {Gaeta}}]{ZhaoOpticaOPTICA2020}%
      \BibitemOpen
      \bibfield  {author} {\bibinfo {author} {\bibfnamefont {Y.}~\bibnamefont {Zhao}}, \bibinfo {author} {\bibfnamefont {X.}~\bibnamefont {Ji}}, \bibinfo {author} {\bibfnamefont {B.~Y.}\ \bibnamefont {Kim}}, \bibinfo {author} {\bibfnamefont {P.~S.}\ \bibnamefont {Donvalkar}}, \bibinfo {author} {\bibfnamefont {J.~K.}\ \bibnamefont {Jang}}, \bibinfo {author} {\bibfnamefont {C.}~\bibnamefont {Joshi}}, \bibinfo {author} {\bibfnamefont {M.}~\bibnamefont {Yu}}, \bibinfo {author} {\bibfnamefont {C.}~\bibnamefont {Joshi}}, \bibinfo {author} {\bibfnamefont {R.~R.}\ \bibnamefont {Domeneguetti}}, \bibinfo {author} {\bibfnamefont {F.~A.~S.}\ \bibnamefont {Barbosa}}, \bibinfo {author} {\bibfnamefont {P.}~\bibnamefont {Nussenzveig}}, \bibinfo {author} {\bibfnamefont {Y.}~\bibnamefont {Okawachi}}, \bibinfo {author} {\bibfnamefont {M.}~\bibnamefont {Lipson}},\ and\ \bibinfo {author} {\bibfnamefont {A.~L.}\ \bibnamefont {Gaeta}},\ }\href {https://doi.org/10.1364/OPTICA.7.000135} {\bibfield  {journal} {\bibinfo  {journal} {Optica}\ }\textbf {\bibinfo {volume} {7}},\ \bibinfo {pages} {135} (\bibinfo {year} {2020})}\BibitemShut {NoStop}%
    \bibitem [{\citenamefont {Stone}\ \emph {et~al.}(2022)\citenamefont {Stone}, \citenamefont {Lu}, \citenamefont {Moille},\ and\ \citenamefont {Srinivasan}}]{stone_efficient_2022}%
      \BibitemOpen
      \bibfield  {author} {\bibinfo {author} {\bibfnamefont {J.~R.}\ \bibnamefont {Stone}}, \bibinfo {author} {\bibfnamefont {X.}~\bibnamefont {Lu}}, \bibinfo {author} {\bibfnamefont {G.}~\bibnamefont {Moille}},\ and\ \bibinfo {author} {\bibfnamefont {K.}~\bibnamefont {Srinivasan}},\ }\href {https://doi.org/10.1063/5.0117691} {\bibfield  {journal} {\bibinfo  {journal} {APL Photonics}\ }\textbf {\bibinfo {volume} {7}},\ \bibinfo {pages} {121301} (\bibinfo {year} {2022})}\BibitemShut {NoStop}%
    \bibitem [{\citenamefont {{Corato-Zanarella}}\ \emph {et~al.}(2024)\citenamefont {{Corato-Zanarella}}, \citenamefont {Ji}, \citenamefont {Mohanty},\ and\ \citenamefont {Lipson}}]{Corato-Zanarella:24}%
      \BibitemOpen
      \bibfield  {author} {\bibinfo {author} {\bibfnamefont {M.}~\bibnamefont {{Corato-Zanarella}}}, \bibinfo {author} {\bibfnamefont {X.}~\bibnamefont {Ji}}, \bibinfo {author} {\bibfnamefont {A.}~\bibnamefont {Mohanty}},\ and\ \bibinfo {author} {\bibfnamefont {M.}~\bibnamefont {Lipson}},\ }\href {https://doi.org/10.1364/OE.505892} {\bibfield  {journal} {\bibinfo  {journal} {Optics Express}\ }\textbf {\bibinfo {volume} {32}},\ \bibinfo {pages} {5718} (\bibinfo {year} {2024})}\BibitemShut {NoStop}%
    \bibitem [{\citenamefont {Moille}\ \emph {et~al.}(2022{\natexlab{b}})\citenamefont {Moille}, \citenamefont {Westly}, \citenamefont {Perez}, \citenamefont {Metzler}, \citenamefont {Simelgor},\ and\ \citenamefont {Srinivasan}}]{MoilleAPLPhotonics2022}%
      \BibitemOpen
      \bibfield  {author} {\bibinfo {author} {\bibfnamefont {G.}~\bibnamefont {Moille}}, \bibinfo {author} {\bibfnamefont {D.}~\bibnamefont {Westly}}, \bibinfo {author} {\bibfnamefont {E.~F.}\ \bibnamefont {Perez}}, \bibinfo {author} {\bibfnamefont {M.}~\bibnamefont {Metzler}}, \bibinfo {author} {\bibfnamefont {G.}~\bibnamefont {Simelgor}},\ and\ \bibinfo {author} {\bibfnamefont {K.}~\bibnamefont {Srinivasan}},\ }\href {https://doi.org/10.1063/5.0127466} {\bibfield  {journal} {\bibinfo  {journal} {APL Photonics}\ }\textbf {\bibinfo {volume} {7}},\ \bibinfo {pages} {126104} (\bibinfo {year} {2022}{\natexlab{b}})}\BibitemShut {NoStop}%
    \end{thebibliography}
%

\end{document}